\documentstyle[prd,aps,preprint,tighten,eqsecnum]{revtex}
\newcommand{\RR}{\hbox{$I$\kern-3.8pt $R$}}
\newcommand{\rr}{\hbox{$\scriptstyle I$\kern-2.4pt $\scriptstyle R$}}
\newcommand{\be}{\begin{equation}}
\newcommand{\ee}{\end{equation}}
\newcommand{\bea}{\begin{eqnarray}}
\newcommand{\eea}{\end{eqnarray}}

\newcommand{\sss}{\scriptscriptstyle}
\newcommand{\diffsigma}{{\rm Diff}\Sigma}
\newcommand{\diffs}{{\rm Diff}\S}

\newcommand{\Sbar}{{\bar S}}

\renewcommand{\S}{{\cal S}}

\newcommand{\boldV}{{\bbox{V}}}

\newcommand{\boldH}{{\bbox{H}}}

\newcommand{\boldN}{{\bbox{N}}}

\newcommand{\bg}{{\bbox{g}}}
\newcommand{\bp}{{\bbox{p}}}
\newcommand{\bTheta}{{\bbox{\Theta}}}
\newcommand{\btheta}{{\bbox{\theta}}}
\newcommand{\bvartheta}{{\bbox{\vartheta}}}
\newcommand{\bPhi}{{\bbox{\Phi}}}
\newcommand{\bphi}{{\bbox{\phi}}}
\newcommand{\bvarphi}{{\bbox{\varphi}}}
\newcommand{\bXi}{{\bbox{\Xi}}}
\newcommand{\bxi}{{\bbox{\xi}}}
\newcommand{\bPi}{{\bbox{\Pi}}}
\newcommand{\bpi}{{\bbox{\pi}}}
\newcommand{\bomega}{{\bbox{\omega}}}
\newcommand{\bchi}{{\bbox{\chi}}}
\begin{document}
\preprint{July 1994} 

\title{On Variational Principles for Gravitating Perfect Fluids}
\author{J. David Brown}
\address{Department of Physics and Department of Mathematics,\\
    North Carolina State University, Raleigh, NC 27695--8202}
\maketitle
\begin{abstract}
The connection is established between two different action principles for
perfect fluids in the context of general relativity. For one of these
actions, $S$, the fluid four--velocity is expressed as a sum of products of
scalar fields and their gradients (the velocity--potential representation).
For the other action, ${\bar S}$, the fluid four--velocity is proportional
to the totally antisymmetric product of gradients of the fluid Lagrangian
coordinates. The relationship between $S$ and ${\bar S}$ is established by
expressing $S$ in Hamiltonian form and identifying certain canonical
coordinates as ignorable. Elimination of these coordinates and their
conjugates yields the action ${\bar S}$. The key step in the analysis is a
point canonical transformation in which all tensor fields on space are
expressed in terms of the Lagrangian coordinate system supplied by the
fluid. The canonical transformation is of interest in its own right. It
can be applied to any physical system that includes a material medium
described by Lagrangian coordinates. The result is a Hamiltonian
description of the system in which the momentum constraint is trivial.
\end{abstract}
\pacs{???}
\section{Introduction}

Many different variational principles have been devised that describe
perfect fluids in the context of general relativity. Some make extensive
use of the Lagrangian coordinate system, the system of spacetime
coordinates defined by the fluid flow lines and the proper time along
those flow lines \cite{Taub}. The use of special coordinates violates
the spirit of general relativity. More importantly, the Lagrangian
time coordinate does not, in general, label spacelike hypersurfaces
and is not suitable for the development of a Hamiltonian version of the
action. Some perfect fluid actions rely on constrained variations in
which the field variables are not freely varied \cite{HandE}. In
practice it is unclear whether or not such an action can be expressed
in Hamiltonian form. Moreover, the usual relationship between the
variational principle and its associated boundary value problem
\cite{Lanczos} is absent because typically the boundary conditions
that follow from the action must be imposed on the parameters of the
constrained field variations rather than on the field variables
themselves.

Some perfect fluid actions employ a set of spacetime scalar
fields as the basic field variables. The fluid four--velocity is then
constructed from combinations of these fields and their
gradients.  This is the velocity--potential or Clebsch potential
\cite{Clebsch} formalism---for general relativistic perfect fluids
this approach was first discussed by Tam and O'Hanlon \cite{TandO'H}
and by Schutz \cite{Schutz}. (See also Ref.~\cite{Sactions}.)
In these papers the theory of Pfaff
forms \cite{Car} was used to minimize the number of velocity potentials.
Thus, the fluid four--velocity $U_\alpha$ was expressed in terms of four
scalar fields,  $\varphi$, $\vartheta$, $W$, and $Z$, as
\be
\mu U_\alpha = -\varphi_{,\alpha} - s\vartheta_{,\alpha} +
W Z_{,\alpha} \ ,\eqnum{1.1}
\ee
where $\mu$ is the fluid chemical potential and $s$ is the entropy
per particle. The velocity
potential formalism was further developed in Ref.~\cite{Brown}
where it was argued that locally the fluid velocity can be
expressed as in Eq.~(1.1), but globally the representation (1.1)
must allow for potentials that are not single--valued
functions on spacelike hypersurfaces. On the other hand, if the
velocity potentials $\varphi$, $\vartheta$, $W$, and $Z$ are required
to be single--valued then the representation (1.1) places a
restriction on the velocity $U_\alpha$. In particular, it can be
shown that the helicity\footnote{Helicity is defined as the
integral over a spacelike hypersurface of $-V\wedge dV$,
where $V:=\mu U$ is the Taub current \cite{TaubV} expressed as
a one--form.} for an isentropic fluid ($s={\rm const}$)
necessarily vanishes when the velocity is given
by expression (1.1) with single--valued potentials.

The shortcomings of the velocity--potential representation (1.1)
are avoided by enlarging the number of potentials and writing
\cite{Brown}
\be
\mu U_\alpha = -\varphi_{,\alpha} - s\vartheta_{,\alpha}
+ W_k {Z^k}_{,\alpha} \ ,\eqnum{1.2}
\ee
where the index $k$ takes the values $1$, $2$, $3$. Thus there are
eight scalar potentials $\varphi$, $\vartheta$, $W_k$, and $Z^k$.
In this case
the fields $Z^k$ can be interpreted as Lagrangian coordinates
for the fluid. That is, the values of the three fields $Z^k(y)$
serve as a set of labels that specify which fluid flow line passes
through the spacetime point $y$. Perfect fluid actions
that are based on the velocity--potential representation (1.2),
with $Z^k$ interpreted as Lagrangian coordinates,
will be denoted by $S$. There are several such actions
corresponding to different functional expressions for the
fluid equation of state \cite{Brown}. These actions
have several features in common with yet another class of perfect
fluid actions first discussed by Carter \cite{Carter} and by
Kijowski {\it et al.\/} \cite{Kijowski}. (See also
Ref.~\cite{Sbaractions}.) These actions will be denoted by
$\Sbar$. They are distinguished by the fact that they are
functionals only of the Lagrangian coordinates $Z^k$. In this
case, the four--velocity is constructed from the antisymmetric
product of gradients of the Lagrangian coordinates,
\be
U^\alpha \sim \epsilon^{\alpha\beta\sigma\rho} {Z^1}_{,\beta}
{Z^2}_{,\sigma} {Z^3}_{,\rho} \ ,\eqnum{1.3}
\ee
where $\epsilon^{\alpha\beta\sigma\rho}$ is obtained by
raising indices on the spacetime volume form
$\epsilon_{\alpha\beta\sigma\rho}$.

The results obtained in Ref.~\cite{Brown} suggest that the actions
$S$ and $\Sbar$ are closely related. The first objective of this
paper is to establish this relationship.

In brief, the connection between the actions $S$ and $\Sbar$
is described as follows. Both of the actions $S$ and $\Sbar$
can be put into standard Hamiltonian form in which the Hamiltonian
is a linear combination of contributions to the Hamiltonian and
momentum constraints. For $S$, the variables $s$ and $W_k$ are
proportional to the momenta conjugate to $\vartheta$ and $Z^k$,
respectively, so the phase space variables are $\varphi$, $\vartheta$,
$Z^k$, and their canonical conjugates. By a point canonical
transformation, the Hamiltonian action $S$ can be expressed in terms
of a new set of canonical variables. The new canonical coordinates
$\bvarphi$ and $\bvartheta$, that in an appropriate sense replace
$\varphi$ and $\vartheta$, are ignorable (cyclic) and can be
eliminated along with their canonical conjugates by Routh's
procedure \cite{Lanczos}. The remaining variables, after
application of the inverse canonical transformation, are
the Lagrangian coordinates $Z^k$ and their conjugates. The
resulting action is the Hamiltonian form of $\Sbar$.

The key step in the analysis is the point canonical transformation.
This transformation is described in terms of a mapping
$Z:\Sigma\to\S$ from the space manifold $\Sigma$ to the
abstract ``fluid space" manifold $\S$ whose points $\zeta\in\S$
are the individual fluid flow lines \cite{Kijowski,Brown}. The
canonical variables $Z^k(x)$ collectively express the
mapping $Z$ in terms of local coordinates $x^a$ on $\Sigma$ and
$\zeta^k$ on $\S$:
\be
Z:\Sigma\to\S \qquad {\rm by} \qquad x^a \mapsto \zeta^k = Z^k(x)
\ .\eqnum{1.4}
\ee
Under the canonical transformation, the canonical coordinate
$Z^k(x)$ is replaced by its inverse $X^a(\zeta)$, which is the
local coordinate expression of the inverse mapping
$X := Z^{-1}$. Whereas $\zeta^k = Z^k(x)$ specifies which flow line
passes through the space point $x^a$, $x^a = X^a(\zeta)$ specifies
the spatial location of the flow line $\zeta^k$.

Under the canonical transformation all other tensor fields are
mapped from space $\Sigma$ to the fluid space $\S$ by $Z$ and its
inverse $X$. Accordingly, the original set of canonical variables
are referred to as the $\Sigma$--variables, while the new set of
canonical variables are referred to as the $\S$--variables. In
particular, the $\Sigma$--variables $\varphi(x)$ and $\vartheta(x)$
are mapped into the $\S$--variables $\bvarphi(\zeta) :=
\varphi(X(\zeta))$ and $\bvartheta(\zeta) := \vartheta(X(\zeta))$.
Their canonical conjugates $\bPi_\varphi(\zeta)$ and
$\bPi_\vartheta(\zeta)$
are the fluid particle number per unit coordinate cell $d^3\zeta$
and the fluid entropy per unit coordinate cell $d^3\zeta$. Since
the particle number and entropy do not change within a flow tube
(defined by a neighborhood of a point $\zeta$ in the fluid space),
the momenta $\bPi_\varphi(\zeta)$ and $\bPi_\vartheta(\zeta)$ are
conserved. In the variational principle the conservation of
$\bPi_\varphi$ and $\bPi_\vartheta$ is a consequence of the
ignorable property of $\bvarphi$ and $\bvartheta$.

The second objective of this paper is to present the details of
the canonical transformation just described. This transformation
can be applied to any system that includes a material medium
described by Lagrangian coordinates $Z^k$. The
result is a canonical formulation of the system in which the
variable $X^a(\zeta)$ and its conjugate $P_a(\zeta)$ are the
only ones that depend on the space manifold $\Sigma$ for their
definitions. Consequently the momentum constraint, which is the
canonical generator of spatial diffeomorphisms, depends only on
the variables $X^a$ and $P_a$. Thus the momentum constraint
expressed in the canonical $\S$--variables is relatively
trivial. This result, expressed in a slightly different form, was
used in Ref.~\cite{dust} to simplify the formal Dirac
quantization of gravity coupled to a pressureless perfect fluid
(dust). In particular, the operator condition on the quantum state
functional that follows from the classical momentum constraint is
trivially solvable. The canonical transformation between
$\Sigma$--variables and $\S$--variables is also utilized in
Ref.~\cite{clocks}, where the action for an elastic medium of
clocks \cite{DeWitt} is analyzed.

Two perfect fluid actions of the velocity--potential type
\cite{Brown} are presented in Section 2a. For one of these
actions, the equation of state is specified by giving $\rho(n,s)$,
the fluid energy density $\rho$ as a function of particle number
density $n$ and entropy per particle $s$. For the other action, the
equation of state is specified by giving $p(\mu,s)$, the fluid
pressure $p$ as a function of chemical potential $\mu$ and
entropy per particle $s$. Symmetries of these actions that
preserve the form of the velocity--potential expression (1.2) are
presented. The Hamiltonian form of the action with $\rho(n,s)$ is
given in Ref.~\cite{Brown}, while the Hamiltonian form
of the action with $p(\mu,s)$ is derived in Appendix~A. These
results are merged in Section 2b. The result is an action $S$
whose associated Hamiltonian is implicitly defined through
the solution to an auxiliary equation that relates the Eulerian
and Lagrangian particle number densities. The distinction between
the choice of $\rho(n,s)$ or $p(\mu,s)$ as the equation of state
arises through the choice of $n$ or $\mu$ as the independent
variable in the auxiliary equation. Conserved Noether charges
associated with the symmetries of the action are displayed. The
discussion in Section 2c addresses the results obtained from an
equation of state in which $n$ and temperature $T$ (or $\mu$ and
$nT$) are the independent thermodynamical variables.

The canonical transformation from the $\Sigma$--variables to
the $\S$--variables is presented in Section 3a, and applied
to the perfect fluid and gravitational actions in Section 3b.
A key mathematical derivation is presented in Appendix B.
The momentum constraint and its role as the generator of space
$\Sigma$ diffeomorphisms is discussed in Section 3c. The
conserved charge that generates fluid space $\S$ diffeomorphisms
is discussed in Section 3d.

In Section~4a the canonical coordinates $\bvarphi$ and $\bvartheta$ are
identified as ignorable variables and the pairs $\bvarphi$,
$\bPi_\varphi$ and $\bvartheta$, $\bPi_\vartheta$ are eliminated by a
straightforward application of Routh's procedure. The resulting action
is denoted $\Sbar$. The symmetries and conserved charges of $\Sbar$
are discussed. In Section 4b the canonical transformation is
reversed for the remaining variables yielding an expression for the
action in terms of $\Sigma$--variables. The Lagrangian form of $\Sbar$
is obtained in Section 4c for the equation of state
$\rho(n,s)$. The fluid four--velocity is expressed as the combination
(1.3) of gradients of the Lagrangian coordinates. In Section 4d a
hybrid action is constructed by eliminating the ignorable pair
$\bvarphi$, $\bPi_\varphi$, and retaining the pair
$\bvartheta$, $\bPi_\vartheta$ as dynamical variables.
This yields the action discussed in Ref.~\cite{Kijowski}.
\section{Perfect Fluid Action with Velocity--Potentials}
\subsection{Lagrangian Actions}

The local expression of the first law of thermodynamics
can be written as
\be
d\rho = \mu\, dn + nT\, ds \ ,\eqnum{2.1}
\ee
where $\rho$ is the energy density, $n$ is the number density, $s$ is
the entropy per particle, $\mu$ is the chemical potential, and $T$ is
the temperature. Pressure is defined by $p := n\mu - \rho$. These
quantities characterize the thermodynamical state of a fluid in
the fluid rest frame.

Equation (2.1) shows that the fluid equation of state can be specified by
the function $\rho(n,s)$. An action principle of the velocity--potential
type that describes a relativistic perfect fluid with equation of state
$\rho(n,s)$ is \cite{Brown}
\bea
\lefteqn{S^{\sss F}[J^\alpha,\varphi,\vartheta,s,W_k,Z^k;
\gamma_{\alpha\beta}] } \qquad\qquad\qquad\nonumber\\
& = & \int d^4y \Bigl\{ -\sqrt{-\gamma} \rho(|J|/\sqrt{-\gamma},s)
+ J^\alpha( \varphi_{,\alpha}+ s\vartheta_{,\alpha}- W_k{Z^k}_{,\alpha})
\Bigr\} \ .\eqnum{2.2}
\eea
Here, $\gamma_{\alpha\beta}$ is the spacetime metric and the magnitude
of $J^\alpha$ is denoted by
$|J|:=\sqrt{-J^\alpha \gamma_{\alpha\beta} J^\beta}$. The variable
$J^\alpha$ is interpreted as the densitized particle number flux
vector; that is, the fluid four--velocity is defined by
\be
U^\alpha := \frac{J^\alpha}{n\sqrt{-\gamma}}  \eqnum{2.3}
\ee
where $n:= |J|/\sqrt{-\gamma}$ is the number density.
The equation of motion obtained by varying the action with respect
to $J^\alpha$ is
\be
\mu U_\alpha = -\varphi_{,\alpha} - s\vartheta_{,\alpha} +
W_k{Z^k}_{,\alpha} \ .\eqnum{2.4}
\ee
This is the velocity--potential expression (1.2).
The potential $\vartheta$ that appears here is the ``thermasy"
\cite{Dantzig}. The equation of motion that follows from varying $s$,
along with the first law (2.1), show that the gradient of $\vartheta$
along the fluid flow lines is the temperature:
\be
\vartheta_{,\alpha} U^\alpha = T \ .\eqnum{2.5}
\ee
The equations of motion obtained by varying $W_k$ imply
\be
{Z^k}_{,\alpha} U^\alpha = 0 \ ,\eqnum{2.6}
\ee
so that the Lagrangian coordinates $Z^k$ are indeed constant along
the fluid flow lines. From these results it follows that the gradient
of the variable $\varphi$ along the flow lines is given by
\be
\varphi_{,\alpha} U^\alpha = \mu - Ts \ ,\eqnum{2.7}
\ee
which is the chemical free energy of the fluid.

The stress--energy--momentum tensor obtained from the action (2.2)
has the form appropriate for a relativistic perfect fluid. Moreover,
it is shown in Ref.~\cite{Brown} that the equations of motion that follow
from this action correctly encode the familiar perfect fluid equations,
namely, the conservation of particle number and entropy per particle along
the fluid flow lines, and the Euler equation relating the fluid acceleration
to the gradient of pressure.

The first law of thermodynamics also can be written in the form
\be
dp = n\, d\mu - nT\, ds \ ,\eqnum{2.8}
\ee
with the energy density defined by $\rho := n\mu - p$. This expression
shows that the fluid equation of state can be specified by the
function $p(\mu,s)$. An action principle of the velocity--potential
type that describes a relativistic perfect fluid with equation of state
$p(\mu,s)$ is \cite{Brown}
\bea
\lefteqn{S^{\sss F}[V^\alpha,\varphi,\vartheta,s,W_k,Z^k;
\gamma_{\alpha\beta}] } \qquad\qquad\nonumber\\
& = & \int d^4y \sqrt{-\gamma} \biggl\{  p(|V|,s) -
\frac{\partial p(|V|,s)}{\partial|V|} \biggl[ |V| - \frac{V^\alpha}{|V|}
\bigl( \varphi_{,\alpha} + s\vartheta_{,\alpha} - W_k{Z^k}_{,\alpha}
\bigr)\biggr]\biggr\} \ .\eqnum{2.9}
\eea
The variable $V^\alpha$ is interpreted as the Taub current \cite{TaubV},
so the fluid four--velocity is defined by
\be
U^\alpha := \frac{V^\alpha}{\mu} \eqnum{2.10}
\ee
where $\mu := |V|$ is the chemical potential. The equation of motion
obtained by varying $V^\alpha$ yields the velocity--potential expression
\be
V_\alpha = -\varphi_{,\alpha} - s\vartheta_{,\alpha} +
W_k{Z^k}_{,\alpha} \ ,\eqnum{2.11}
\ee
which coincides with Eq.~(2.4). The velocity potentials that appear in the
action (2.9) have the same physical interpretation as those that appear
in the action (2.2).

As shown in Ref.~\cite{Brown}, the action (2.9) yields the correct
stress--energy--momentum tensor and the correct perfect fluid equations
of motion. A simpler version of this action is obtained by replacing
$V^\alpha$ with the solution (2.11) of the $V^\alpha$ equation of motion.
This yields $S^{\sss F} = \int d^4y \sqrt{-\gamma} p(|V|,s)$ where
$V^\alpha$ is
shorthand notation for the right--hand side of Eq.~(2.11). Both versions
lead to the same action in Hamiltonian form; the derivation in Appendix A
uses the action in Eq.~(2.9) as a starting point.

The actions (2.2) and (2.9) are invariant under certain symmetry
transformations that preserve the form of the velocity potential expression
on the right--hand sides of Eqs.~(2.4), (2.11). These symmetries
include \cite{Brown,dust} invariance with respect to a change of Lagrangian
coordinate labels,
\be
Z^k = \bXi^k(Z') \ ,\qquad W'_{k} =
\frac{\partial\bXi^\ell(Z')}{\partial{Z'}^k} W_\ell
\ ,\eqnum{2.12a}
\ee
invariance with respect to a deformation or ``tilt" of the
constant--$\varphi$ surfaces,
\be
\varphi' = \varphi + \bPhi(Z) \ ,\qquad W'_{k} = W_k +
\frac{\partial\bPhi(Z)}{\partial Z^k} \ ,\eqnum{2.12b}
\ee
and invariance with respect to a deformation of the
constant--$\vartheta$ surfaces,
\be
\vartheta' = \vartheta + \bTheta(Z) \ ,\qquad W'_{k} = W_k +
s\frac{\partial\bTheta(Z)}{\partial Z^k} \ .\eqnum{2.12c}
\ee
The invariances (2.12a--c) give rise to conserved Noether currents
and charges. For the invariance (2.12a), the Noether current is related
to the fluid momentum in the directions orthogonal to
$\varphi_{,\alpha} +  s\vartheta_{,\alpha}$ \cite{dust}.
The Noether currents associated with the invariances (2.12b) and (2.12c)
are expressions of the conserved particle number current and the
conserved entropy current \cite{Brown}.
\subsection{Hamiltonian Action}

The Hamiltonian form of the action (2.2) is derived in Ref.~\cite{Brown},
and the Hamiltonian form of the action (2.9) is derived in Appendix A.
These results are summarized as follows. The Hamiltonian action is
\bea
\lefteqn{S^{\sss F}[\varphi,\Pi_\varphi,\vartheta,\Pi_\vartheta,
Z^k,P_k;g_{ab},N^\perp,N^a] }\qquad\qquad\qquad\qquad\nonumber\\
& = & \int_{\rr} dt \int_\Sigma d^3x \Bigl( \Pi_\varphi {\dot\varphi}
+ \Pi_\vartheta {\dot\vartheta} + P_k{\dot Z}^k - N^\perp H^{\sss F}_\perp
- N^a H^{\sss F}_a \Bigr) \ ,\eqnum{2.13}
\eea
where $g_{ab}$ is the metric on space $\Sigma$ and $N^\perp$ and $N^a$
are the lapse function and shift vector, respectively. The fluid
contributions to the momentum and Hamiltonian constraints are
\bea
H^{\sss F}_a  & = & \Pi_\varphi \varphi_{,a} + \Pi_\vartheta \vartheta_{,a}
+ P_k {Z^k}_{,a} \ ,\eqnum{2.14a} \\
H^{\sss F}_\perp & = & \sqrt{\mu^2 \Pi_\varphi^2 + H^{\sss F}_a g^{ab}
H^{\sss F}_b } - \sqrt{g} p \ .\eqnum{2.14b}
\eea
In the expression for $H^{\sss F}_\perp$, $\mu$ and
$p$ are determined as functions of the canonical variables as follows.
First, choose an equation of state, either
$\rho(n,s)$ or $p(\mu,s)$. For the case $\rho(n,s)$, define
\be
\mu(n,s) := \frac{\partial\rho(n,s)}{\partial n} \ ,\qquad p(n,s)
:= n\frac{\partial\rho(n,s)}{\partial n} -\rho(n,s) \ ;\eqnum{2.15a}
\ee
for the case $p(\mu,s)$, define
\be
n(\mu,s) := \frac{\partial p(\mu,s)}{\partial\mu} \ .\eqnum{2.15b}
\ee
Next solve the equation
\be
\frac{\Pi_\varphi}{\sqrt{g}} = n \sqrt{ 1 + \frac{H^{\sss F}_a g^{ab}
H^{\sss F}_b}{ \mu^2 \Pi_\varphi^2} } \eqnum{2.16}
\ee
for either $n$ or $\mu$ (depending on the equation of state used) as
a function of $\Pi_\varphi$, $H^{\sss F}_a$, $g_{ab}$, and $s$.
Now make the identification
\be
s := \frac{\Pi_\vartheta}{\Pi_\varphi} \eqnum{2.17}
\ee
for the entropy per particle $s$. In this way either $n$ or $\mu$
(depending on the equation of state used) is
determined as a function of $\Pi_\varphi$, $\Pi_\vartheta$,
$H^{\sss F}_a$, and $g_{ab}$. This result along with the equation of
state and the definitions (2.15a) or (2.15b) determine $\mu$ and $p$,
and subsequently $H^{\sss F}_\perp$, as functions of $\Pi_\varphi$,
$\Pi_\vartheta$, $H^{\sss F}_a$, and $g_{ab}$.

The momentum $\Pi_\varphi$ is the particle number on $\Sigma$ per
unit coordinate cell $d^3x$. The auxiliary equation (2.16) relates
the Eulerian number density $\Pi_\varphi/\sqrt{g}$ (the number
density per unit proper volume) to the Lagrangian number density $n$.
The square root factor in Eq.~(2.16) is just the relativistic gamma
factor $\sqrt{1 + U_a g^{ab} U_b}$, where $U_a =
- H^{\sss F}_a/(\mu\Pi_\varphi)$ are the spatial components
of the fluid four--velocity.

The Noether charges associated with the symmetries (2.12a) are
\be
Q[\vec\bxi] = -\int_\Sigma d^3x \, P_k(x) \bxi^k(Z(x))
\ ,\qquad
\bxi^k(Z) := \left.\frac{\partial\bXi^k(Z,\sigma)}{\partial\sigma}
\right|_{\sigma = 0} \ ,\eqnum{2.18}
\ee
where $\bXi(Z,\sigma)$ denotes a one--parameter subgroup of
transformations (2.12a). Likewise, the Noether charges associated
with the symmetries (2.12b) are
\be
Q[\bphi] = \int_\Sigma d^3x\,  \Pi_\varphi(x) \bphi(Z(x))
\ ,\qquad
\bphi(Z) := \left.\frac{\partial\bPhi(Z,\sigma)}{\partial\sigma}
\right|_{\sigma = 0} \ ,\eqnum{2.19}
\ee
and the charges associated with the symmetries (2.12c) are
\be
Q[\btheta] = \int_\Sigma d^3x\,  \Pi_\vartheta(x) \btheta(Z(x))
\ ,\qquad
\btheta(Z) := \left.\frac{\partial\bTheta(Z,\sigma)}{\partial\sigma}
\right|_{\sigma = 0} \ .\eqnum{2.20}
\ee
The charges (2.18--20) generate the infinitesimal versions of the
symmetries (2.12) through the Poisson brackets. (The derivation of
the Hamiltonian action shows that
$W_k = - P_k/\Pi_\varphi$.) In particular, the charge (2.18) is the
canonical generator of $\diffs$, fluid space diffeomorphisms, where
the fluid space vector field $\bxi^k(\zeta)$ is an element of the
Lie algebra of $\diffs$ \cite{dust}.
\subsection{Other Equations of State}

For the sake of completeness, it is appropriate to finish this
section with a comment on other possible expressions for the
equation of state. Observe that the first law as expressed in
Eq.~(2.8) is obtained from the first law (2.1)
by a Legendre transformation in $n$ and $\mu$. A Legendre
transformation can be performed in $T$ and $s$ instead, which
leads to the first law in the form
\be
d(na) = (\mu - Ts) dn - ns\, dT \ .\eqnum{2.21}
\ee
Here $a = \rho/n -Ts$ is the physical free energy. This result
suggests the use of an equation of state of the form $a(n,T)$.
The perfect fluid action with this equation of state is obtained
by replacing $\rho(n,s)$ in Eq.~(2.2) with $na(n,T) + nTs$
\cite{Brown}. The temperature $T$ is treated as a new variable;
the equation of motion obtained by varying $T$ reproduces the
definition of $s$ from the first law, namely,
\be
s =  - \frac{\partial a(n,T)}{\partial T} \ .\eqnum{2.22}
\ee
The Hamiltonian form of the resulting action coincides with the
Hamiltonian action (2.13) with $na(n,T) + nTs$ replacing
$\rho(n,s)$. However, this action is not truly in Hamiltonian
form, since the extra variable $T$ appears. $T$ can be eliminated
by inverting Eq.~(2.22) for $T(n,s)$ and substituting the result
back into the Hamiltonian.\footnote{One can attempt to retain $T$
as a canonical variable by defining its conjugate momentum $P_T$.
However, this leads to a primary constraint $P_T=0$ and a secondary
constraint which is Eq.~(2.22) written in terms of the canonical
variables. This pair of constraints is second class, and its
elimination is equivalent to the result obtained by solving
Eq.~(2.22) for $T$ and substituting the solution back into the
Hamiltonian.} Through this
process the equation of state $a(n,T)$ is replaced by the equation
of state $\rho(n,s)$, and the Hamiltonian action is converted
back into the action (2.13) with equation of state $\rho(n,s)$.
In a similar manner, a Legendre transformation in $nT$ and $s$
can be applied to the first law (2.8), leading to an equation
of state in which $p+ nTs$ is given as a function of $\mu$ and $nT$.
The corresponding action can be obtained from Eq.~(2.9).
In passing to the Hamiltonian form of this action the variable $nT$
is eliminated, resulting in the action (2.13) with equation of state
$p(\mu,s)$.
\section{Canonical $\S$--Variables}
\subsection{The Canonical Transformation}

The main points of this section are best illustrated with the perfect fluid
coupled to other fields. Thus, consider a perfect fluid coupled to the
gravitational field. The action for the combined system is obtained
by adding the gravitational action to the fluid action $S^{\sss F}$. The
result, in Hamiltonian form, is
\bea
\lefteqn{S[\varphi,\Pi_\varphi,\vartheta,\Pi_\vartheta,
Z^k,P_k,g_{ab},p^{ab},N^\perp,N^a] }\qquad\qquad\nonumber\\
& = & \int_{\rr} dt \int_\Sigma d^3x \Bigl( \Pi_\varphi {\dot\varphi}
+ \Pi_\vartheta {\dot\vartheta} + P_k{\dot Z}^k + p^{ab} {\dot g}_{ab}
- N^\perp H_\perp - N^a H_a \Bigr) \ ,\eqnum{3.1}
\eea
where $p^{ab}$ is the momentum conjugate to the metric $g_{ab}$.
The Hamiltonian and momentum constraints are
\bea
H_\perp = H^{\sss G}_\perp + H^{\sss F}_\perp \ ,\eqnum{3.2a}\\
H_a = H^{\sss G}_a + H^{\sss F}_a \ ,\eqnum{3.2b}
\eea
where $H^{\sss F}_\perp$ and $H^{\sss F}_a$ are the fluid
contributions (2.14) and $H^{\sss G}_\perp$ and $H^{\sss G}_a$
are the familiar gravitational field contributions.

The canonical variables that appear in the action (3.1)
are the pairs $Z^k(x)$ and $P_k(x)$, $\varphi(x)$ and $\Pi_\varphi(x)$,
$\vartheta(x)$ and $\Pi_\vartheta(x)$, and $g_{ab}(x)$ and $p^{ab}(x)$.
These are the canonical $\Sigma$--variables. Note that
$\varphi(x)$, $\Pi_\varphi(x)$, $\vartheta(x)$,
$\Pi_\vartheta(x)$, $g_{ab}(x)$, and $p^{ab}(x)$ are (time
$t$--dependent) tensor fields on $\Sigma$. The variable $Z^k(x)$ is the
local coordinate chart expression of a ($t$--dependent)
mapping $Z:\Sigma\to\S$ from the space manifold $\Sigma$ to the fluid
space $\S$. Its conjugate $P_k(x)$ is the coordinate expression of
a ($t$--dependent) mapping from $\Sigma$ to $T^*\S$, with density
weight $1$ on $\Sigma$.
In this section an alternative set of canonical variables, the
$\S$--variables, is described. The $\S$--variables utilize the fact
that on $\Sigma$ the fluid particles naturally define a preferred set
of spatial coordinates.

First, introduce the $t$--dependent mapping $X$, which is the inverse
of $Z$:
\begin{equation} X:\S\to\Sigma \ ,\qquad X := Z^{-1}
        \ .\eqnum{3.3}
\end{equation}
The mapping $X$ specifies the spatial coordinates of the fluid flow lines;
that is, $x^a = X^a(\zeta)$ is the spatial location of the fluid flow line
whose Lagrangian coordinate label is $\zeta^k$. The mappings $Z$ and $X$
induce mappings of tensor fields from $\Sigma$ to $\S$. In particular,
for the scalars $\varphi$, $\vartheta$ and the covariant metric
tensor $g_{ab}$, the corresponding tensors on $\S$ are the pullbacks
of those fields by $X$:
\bea
\bvarphi & := & X^* \varphi  = \varphi\circ X  \ ,\eqnum{3.4a}\\
\bvartheta & := & X^* \vartheta  = \vartheta\circ X  \ ,\eqnum{3.4b}\\
\bg  & := & X^* g \ .\eqnum{3.4c}
\eea
In local coordinates, these definitions become
\bea
\bvarphi(\zeta) & = & \varphi(X(\zeta)) \ ,\eqnum{3.5a}\\
\bvartheta(\zeta) & = & \vartheta(X(\zeta)) \ ,\eqnum{3.5b}\\
\bg_{k\ell}(\zeta) & = & {X^a}_{,k}(\zeta) {X^b}_{,\ell}(\zeta)
     \, g_{ab}(X(\zeta)) \ .\eqnum{3.5c}
\eea
The variables $\bg_{k\ell}(\zeta)$ are the components of the metric
tensor
\be
ds^2 = \bg_{k\ell}(\zeta)\, d\zeta^k d\zeta^\ell
   \eqnum{3.6}
\ee
on the fluid space $\S$. That is, $\bg_{k\ell}(\zeta)$ measures the proper
distance $ds$ in space $\Sigma$ between neighboring fluid particles with
Lagrangian coordinate labels $\zeta^k$ and $\zeta^k + d\zeta^k$.

The canonical transformation from the $\Sigma$--variables to the
$\S$--variables is a point transformation of the canonical coordinates
defined by
\be
\Bigl( Z^k(x),\  \varphi(x),\  \vartheta(x),\  g_{ab}(x) \Bigr)
\mapsto \Bigl( X^a(\zeta),\  \bvarphi(\zeta),\  \bvartheta(\zeta),\
\bg_{k\ell}(\zeta) \Bigr)  \ .\eqnum{3.7}
\ee
The momenta conjugate to the new coordinates $X^a(\zeta)$,
$\bvarphi(\zeta)$, $\bvartheta(\zeta)$
and $\bg_{k\ell}(\zeta)$ are identified by imposing the invariance of
the canonical one--form, or equivalently, the invariance of
\be
\int_\Sigma d^3x \Bigl\{ P_k(x) {\dot Z}^k(x)  + \Pi_\varphi(x)
{\dot\varphi}(x)  + \Pi_\vartheta(x) {\dot\vartheta}(x) +
p^{ab}(x) {\dot g}_{ab}(x) \Bigr\} \ .\eqnum{3.8}
\ee
Note that $x^a$ and $\zeta^k$ play the role of continuous
indices that label the different canonical coordinates and momenta.

As mappings between the space manifold $\Sigma$ and the fluid space
manifold $\S$, $Z^k(x)$ and $X^a(\zeta)$ play a special role in the
canonical transformation. Consequently a separate analysis will
be presented for the terms $P_k {\dot Z}^k$ in Eq.~(3.8). The
remaining terms in Eq.~(3.8) are constructed from tensor fields on
$\Sigma$, and their analysis is presented in Appendix B.

To begin, use the fact that $Z\circ X$
is the identity on $\S$ to express ${\dot Z}^k(x)$ in
terms of ${\dot X}^a(\zeta)$. In local coordinates, we have
\be
Z^k(X(\zeta)) = \zeta^k \ ,\eqnum{3.9}
\ee
and differentiation with respect to $t$ yields
\be
{\dot Z}^k(x) \Bigr|_{x=X(\zeta)} = - \biggl( {Z^k}_{,a}(x)
\Bigr|_{x=X(\zeta)} \biggr) {\dot X}^a(\zeta)  \ .\eqnum{3.10}
\ee
This equation can be written more compactly as
${\dot Z}^k(X(\zeta)) = - {Z^k}_{,a}(X(\zeta)) {\dot X}^a(\zeta)$. Now
use this result along with a change of integration variables
$x^a = X^a(\zeta)$ in the first term of expression (3.8) to obtain
\bea
\int_\Sigma d^3x \, P_k(x) {\dot Z}^k(x) & = &
\int_\S d^3\zeta \left|\frac{\partial X(\zeta)}{\partial\zeta}\right|
P_k(X(\zeta))\, {\dot Z}^k(X(\zeta))  \eqnum{3.11a}\\
& = & - \int_\S d^3\zeta \left|\frac{\partial X(\zeta)}
{\partial\zeta}\right| P_k(X(\zeta)) \,{Z^k}_{,a}(X(\zeta))
\,{\dot X}^a(\zeta) \ .\eqnum{3.11b}
\eea
Here, $|\partial X(\zeta) / \partial\zeta|$ is the Jacobian of the change
of variables.\footnote{ The change of integration variables
$x^a = X^a(\zeta)$, which involves the
new canonical coordinate $X^a(\zeta)$, is described more precisely
as follows. The integrand on the left--hand side of Eq.~(3.11a) is a
three--form on  $\Sigma$, call it $\omega$. Its counterpart on $\S$ is
the three--form $\bomega := X^*\omega$ obtained by the pullback mapping
$X^*$. In terms of local coordinates, $\bomega_{k\ell m}(\zeta) =
{X^a}_{,k}(\zeta) {X^b}_{,\ell}(\zeta) {X^c}_{,m}(\zeta)\,
\omega_{abc}(X(\zeta))$.
Equation (3.11a) just expresses the fundamental identity
$\int_\Sigma \omega = \int_\S \bomega$
in terms of local coordinates. Thus, the change of variables
$x^a = X^a(\zeta)$ in Eq.~(3.11)
arises because the integrand (the three--form) is being mapped
from $\Sigma$ to $\S$. This is the way in which the continuous index set
$x^a$ is changed to the continuous index set $\zeta^k$.}

The remaining terms in Eq.~(3.8) are rewritten by making use of the
following result, which is derived in Appendix B. Consider an arbitrary
$t$--dependent tensor field $\chi$ on $\Sigma$, and its conjugate
tensor density $\pi$. (Indices on $\chi$ and $\pi$ are suppressed, and
summation of indices is implied in expressions such as
$\pi\dot\chi$.) Also let $H_a(x;\pi,\chi]$ denote the contribution from
$\chi$ and $\pi$ to the momentum constraint. The key result
from Appendix B is
\be
\int_{\Sigma} d^3x \,{\pi}(x) \dot{\chi}(x) =
\int_\S d^3\zeta \biggl( \bpi(\zeta) {\dot\bchi}(\zeta) -
\boldH_k(\zeta;\bpi,\bchi]\, {Z^k}_{,a}(X(\zeta)) \,{\dot X}^a(\zeta)
\biggr) \ ,\eqnum{3.12}
\ee
where $\bchi(\zeta)$, $\bpi(\zeta)$, and $\boldH_k(\zeta;\bpi,\bchi]$ are
the tensors (tensor densities) on $\S$ that correspond to $\chi(x)$,
$\pi(x)$, and $H_a(x;\pi,\chi]$, respectively.

The results (3.11) and (3.12) can be applied to expression (3.8) with
the result
\bea
\lefteqn{ \int_\Sigma d^3x \Bigl\{ P_k(x) {\dot Z}^k(x)  +
\Pi_\varphi(x) {\dot\varphi}(x) + \Pi_\vartheta(x) {\dot\vartheta}(x)
+ p^{ab}(x) {\dot g}_{ab}(x) \Bigr\} } \quad\nonumber\\
& & = \int_\S d^3\zeta \Bigl\{ -\boldH_k(\zeta) {Z^k}_{,a}(X(\zeta))
{\dot X}^a(\zeta) + \bPi_\varphi(\zeta){\dot\bvarphi}(\zeta)
+ \bPi_\vartheta(\zeta){\dot\bvartheta}(\zeta) + \bp^{k\ell}(\zeta)
{\dot\bg}_{k\ell}(\zeta) \Bigr\} \ .\eqnum{3.13}
\eea
Here, $\bPi_\varphi(\zeta)$, $\bPi_\vartheta(\zeta)$, and
$\bp^{k\ell}(\zeta)$ are the fluid space tensor
densities that correspond to the fields $\Pi_\varphi(x)$,
$\Pi_\vartheta(x)$, and $p^{ab}(x)$:
\bea
\bPi_\varphi(\zeta) & := & \left|\frac{\partial X(\zeta)}
{\partial\zeta}\right| \Pi_\varphi(X(\zeta)) \ ,\eqnum{3.14a}\\
\bPi_\vartheta(\zeta) & := & \left|\frac{\partial X(\zeta)}
{\partial\zeta}\right| \Pi_\vartheta(X(\zeta)) \ ,\eqnum{3.14b}\\
\bp^{k\ell}(\zeta) & := & \left|\frac{\partial X(\zeta)}
{\partial\zeta}\right| {Z^k}_{,a}(X(\zeta)) {Z^\ell}_{,b}(X(\zeta))
\, p^{ab}(X(\zeta)) \ .\eqnum{3.14c}
\eea
Also, $\boldH_k(\zeta)$ is the fluid space covariant vector density
obtained by mapping the full momentum constraint $H_a(x)$ from $\Sigma$
to $\S$:
\be
\boldH_k(\zeta) := \left|\frac{\partial X(\zeta)}
{\partial\zeta}\right| {X^a}_{,k}(\zeta) H_a(X(\zeta))
     \ .\eqnum{3.15}
\ee
Equation~(3.13) shows that the momentum conjugate to $\bvarphi(\zeta)$
is $\bPi_\varphi(\zeta)$, which is the fluid particle number on $\S$
per unit coordinate cell $d^3\zeta$. Likewise the momentum conjugate to
$\bvartheta(\zeta)$ is $\bPi_\vartheta(\zeta)$, the fluid entropy on
$\S$ per unit coordinate cell $d^3\zeta$. The momentum conjugate
to $\bg_{k\ell}(\zeta)$ is $\bp^{k\ell}(\zeta)$, the expression of the
gravitational momentum in the fluid coordinate system $\zeta^k$. Finally,
the momentum conjugate to $X^a(\zeta)$ is defined by
\begin{equation} P_a(\zeta) := - \boldH_k(\zeta) \,{Z^k}_{,a}(X(\zeta))
   = - \left|\frac{\partial X(\zeta)}{\partial\zeta}\right| H_a(X(\zeta))
   \ ,\eqnum{3.16}\end{equation}
and is proportional to the momentum constraint (3.2b) for the system.

\subsection{The Action as a Functional of $\S$--Variables}

With the results (3.13)--(3.16), the action (3.1) becomes
\bea
\lefteqn{ S[X^a,P_a,\bvarphi,\bPi_\varphi,\bvartheta,\bPi_\vartheta,
\bg_{k\ell},\bp^{k\ell},N^\perp,N^a] } \qquad\quad\nonumber\\
& & = \int_{\rr}dt \int_\S d^3\zeta \biggl\{ P_a(\zeta){\dot X}^a(\zeta)
+ \bPi_\varphi(\zeta){\dot\bvarphi}(\zeta)
+ \bPi_\vartheta(\zeta){\dot\bvartheta}(\zeta) + \bp^{k\ell}(\zeta)
{\dot\bg}_{k\ell}(\zeta) \nonumber\\
& &\qquad\qquad\qquad\qquad\qquad\qquad\qquad - N^\perp(X(\zeta))
    \boldH_\perp(\zeta)  + N^a(X(\zeta)) P_a(\zeta) \biggr\}
    \ ,\eqnum{3.17}
\eea
where the Hamiltonian constraint (3.2a) is expressed in the fluid
space $\S$ as
\bea
\boldH_\perp(\zeta) & := & \left|\frac{\partial X(\zeta)}
{\partial\zeta}\right| H_\perp(X(\zeta)) \ ,\nonumber\\
& = & \boldH_\perp^{\sss G}  + \sqrt{ \mu^2 \bPi_\varphi^2
+ \boldH_k^{\sss F} \bg^{k\ell} \boldH_\ell^{\sss F} } -
\sqrt{\bg} p \ .\eqnum{3.18}
\eea
The gravitational part $\boldH_\perp^{\sss G}(\zeta)$ of the
Hamiltonian constraint is obtained from $H_\perp^{\sss G}(x)$ by
replacing the gravitational $\Sigma$--variables $g_{ab}(x)$,
$p^{ab}(x)$ with the corresponding $\S$--variables $\bg_{k\ell}(\zeta)$,
$\bp^{k\ell}(\zeta)$.  For the perfect fluid part of the Hamiltonian
constraint (3.18), $\mu$ and $p$ are determined as before: First
choose an equation of state $\rho(n,s)$ or $p(\mu,s)$ and define
the relevant thermodynamical variables as in Eqs.~(2.15). Next,
solve the equation
\be
\frac{\bPi_\varphi}{\sqrt{\bg}} = n \sqrt{ 1 +
\frac{\boldH_k^{\sss F} \bg^{k\ell} \boldH_\ell^{\sss F}}
{\mu^2\bPi_\varphi^2} } \eqnum{3.19}
\ee
for $n$ or $\mu$ as a function of $\bPi_\varphi(\zeta)$,
$\boldH_k^{\sss F}(\zeta)$, $\bg_{k\ell}(\zeta)$, and $s$. With
the identification $s := \bPi_\vartheta/\bPi_\varphi$, then $\mu$
and $p$ are determined as functions of $\bPi_\varphi(\zeta)$,
$\bPi_\vartheta(\zeta)$, $\boldH_k^{\sss F}(\zeta)$, and
$\bg_{k\ell}(\zeta)$. The fluid space density $\boldH_k^{\sss F}(\zeta)$
that appears here and explicitly in the Hamiltonian constraint (3.18)
is the fluid contribution to the momentum constraint, mapped to $\S$.
It is expressed as a function of the canonical $\S$--variables through
the solution of Eq.~(3.16) with the identification
$\boldH_k := \boldH_k^{\sss F} + \boldH_k^{\sss G}$. This yields
\be
\boldH_k^{\sss F}(\zeta) = - P_a(\zeta) {X^a}_{,k}(\zeta) -
\boldH_k^{\sss G}(\zeta) \ ,\eqnum{3.20}
\ee
where $\boldH_k^{\sss G}(\zeta)$ is the gravitational contribution to
the momentum constraint, mapped to $\S$. Thus, $\boldH_k^{\sss G}(\zeta)$
depends only on the gravitational canonical variables
$\bg_{k\ell}(\zeta)$ and $\bpi^{k\ell}(\zeta)$.

We can make a further change of variables in the action by defining fluid
space tensors that correspond to the lapse function and shift vector:
\bea
\boldN^\perp(\zeta) & := & N^\perp(X(\zeta))
\ ,\eqnum{3.21a}\\
\boldN^k(\zeta) & := & N^a(X(\zeta)) {Z^k}_{,a}(X(\zeta))
\ .\eqnum{3.21b}
\eea
The action then becomes
\bea
\lefteqn{ S[X^a,P_a,\bvarphi,\bPi_\varphi,\bvartheta,\bPi_\vartheta,
\bg_{k\ell},\bp^{k\ell},\boldN^\perp,\boldN^k] } \qquad\qquad\nonumber\\
& & = \int_{\rr}dt \int_\S d^3\zeta \biggl\{ P_a(\zeta){\dot X}^a(\zeta)
+ \bPi_\varphi(\zeta){\dot\bvarphi}(\zeta)
+ \bPi_\vartheta(\zeta){\dot\bvartheta}(\zeta) + \bp^{k\ell}(\zeta)
{\dot\bg}_{k\ell}(\zeta) \nonumber\\
& &\qquad\qquad\qquad\qquad\qquad\qquad\qquad\qquad\quad
- \boldN^\perp(\zeta) \boldH_\perp(\zeta)  - \boldN^k(\zeta)
\boldH_k(\zeta) \biggr\} \ ,\eqnum{3.22}
\eea
where the momentum constraint
\be
\boldH_k(\zeta) :=  - P_a(\zeta) {X^a}_{,k}(\zeta)  = 0 \eqnum{3.23}
\ee
from Eq.~(3.16) depends only on the
fluid coordinates $X^a$ and their conjugates $P_a$.

The Hamiltonian constraint $H_\perp(x)$ that appears in the action
(3.1) is smeared with a spatial scalar $N^\perp(x)$. This defines
a canonical generator,
\be
H[N^\perp] := \int_\Sigma d^3x \, N^\perp(x) H_\perp(x)
 = \int_\S d^3\zeta\, N^\perp(X(\zeta)) \boldH_\perp(\zeta)
\ ,\eqnum{3.24}
\ee
which, as shown here, can be written either in terms of the canonical
$\Sigma$--variables or the canonical $\S$--variables. On the other
hand, the Hamiltonian constraint that appears in the action
(3.22) is smeared with a fluid space scalar $\boldN^\perp(\zeta)$.
The corresponding canonical generator is
\be
H[\boldN^\perp] := \int_\S d^3\zeta\, \boldN^\perp(\zeta)
\boldH_\perp(\zeta) = \int_\Sigma d^3x \,
\boldN^\perp(Z(x)) H_\perp(x)  \ .\eqnum{3.25}
\ee
Now, the Poisson bracket of any canonical variable $F$ with the
smeared constraint $H[N^\perp]$ equals
\bea
\{ F , H[N^\perp] \} & = & \int_\S d^3\zeta \biggl(
N^\perp(X(\zeta)) \{ F , \boldH_\perp(\zeta) \} +
\{ F , N^\perp(X(\zeta)) \} \boldH_\perp(\zeta) \biggr) \nonumber\\
& = & \{ F , H[\boldN^\perp] \} + \int_\S d^3\zeta \,
\{ F , N^\perp(X(\zeta)) \}   \boldH_\perp(\zeta) \ ,\eqnum{3.26}
\eea
where $\boldN^\perp$ and $N^\perp$ are related as in Eq.~(3.21a).
This result shows that the brackets $\{ F,H[\boldN^\perp]\}$ and
$\{ F,H[N^\perp]\}$ are equal modulo the constraints $H_\perp(x) =
0 = \boldH_\perp(\zeta)$. Thus, on the constraint surface, the
Poisson bracket of $F$ with $H[\boldN^\perp]$ gives
the change in $F$ under a displacement of the hypersurface $\Sigma$
by a proper time $\boldN^\perp(Z(x))$ in the direction orthogonal
to the hypersurface.
\subsection{Momentum Constraint and $\diffsigma$}

The momentum constraint $H_a(x)$ smeared with an externally prescribed
vector field $N^a(x)$ on $\Sigma$ can be written either in terms of the
canonical $\Sigma$--variables or the canonical $\S$--variables:
\be
H[\vec N] := \int_\Sigma d^3x \, N^a(x) H_a(x)
= - \int_\S d^3\zeta \, N^a(X(\zeta)) P_a(\zeta) \ .\eqnum{3.27}
\ee
$H[\vec N]$ is the generator of $\diffsigma$ through the Poisson brackets.
That is, for any canonical variable $F$, the Poisson bracket
\be
\{ F,H[\vec N] \}  \eqnum{3.28}
\ee
gives the change in $F$ due to an infinitesimal diffeomorphism of
$\Sigma$ generated by the vector field $\vec N$.
If $F$ is expressed solely in terms of $\Sigma$--variables, then the Poisson
bracket is most conveniently computed using the $\Sigma$--variables and
the first integral in Eq.~(3.27). If $F$ is expressed solely in terms of
$\S$--variables, then the Poisson bracket is most conveniently computed
using the $\S$--variables and the second integral in Eq.~(3.27). If $F$ is
a tensor field on $\Sigma$, then the Poisson bracket (3.28) yields the
Lie derivative $\pounds_{\vec N} F$.

Observe that the $\S$--variables $\bvarphi$, $\bPi_\varphi$, $\bvartheta$,
$\bPi_\vartheta$, $\bg$, and $\bp$ have
vanishing Poisson brackets with $H[\vec N]$. This simply reflects the
fact that they are tensor fields on the fluid space $\S$ and are
invariant under $\diffsigma$. Among the canonical $\S$--variables,
only $X^a$ and $P_a$ have nontrivial transformation properties under
$\diffsigma$:
\bea
\{ X^a(\zeta) , H[\vec N] \} & = & - N^a(X(\zeta)) \ ,\eqnum{3.29a}\\
\{ P_a(\zeta) , H[\vec N] \} & = & P_b(\zeta) {N^b}_{,a}(X(\zeta))
\ .\eqnum{3.29b}
\eea
The transformation (3.29a) can be verified as follows. Let
${\cal N}_\sigma :\Sigma\to\Sigma$ denote the one--parameter family
of diffeomorphisms of $\Sigma$ with generator
\begin{equation} {\vec N} := \frac{d{\cal N}_\sigma}{d\sigma}
    \biggr|_{\sigma=0} \ .\eqnum{3.30}\end{equation}
Also let ${\cal N}_\sigma^*$ generically denote the action of
${\cal N}_\sigma$ on the canonical field variables. For the fields
$Z^k(x)$, viewed as scalars on $\Sigma$, the action of
${\cal N}_\sigma$ is given by the pullback
${\cal N}_\sigma^* Z^k = Z^k \circ {\cal N}_\sigma$. The associated
infinitesimal diffeomorphism is then
\be
\frac{d}{d\sigma} \Bigl( {\cal N}^*_\sigma Z^k(x) \Bigr) \Bigr|_{\sigma=0}
=  {Z^k}_{,a}(x) N^a(x) = \{ Z^k(x), H[\vec N] \} \ .\eqnum{3.31}
\ee
Correspondingly, the action of $\diffsigma$ on the mapping $Z$ is defined by
${\cal N}^*_\sigma Z  = Z \circ {\cal N}_\sigma$. Since $X$ is the inverse of
$Z$, the action of $\diffsigma$ on $X$ is given by
${\cal N}^*_\sigma X = {\cal N}_\sigma^{-1}\circ X$. For the associated
infinitesimal diffeomorphism, this yields
\be
\frac{d}{d\sigma} \Bigl( {\cal N}^*_\sigma X^a(\zeta) \Bigr)
\Bigr|_{\sigma=0} = - {N^a}(X(\zeta)) \ ,\eqnum{3.32}
\ee
which agrees with Eq.~(3.29a).

The transformation (3.29b) can be checked by recalling that $P_a(\zeta)$
is a mapping from $\S$ to $T^*\Sigma$ with density weight 1 on $\S$.
Thus,  $P_a(\zeta) {X^a}_{,k}(\zeta)$ is a covariant vector density on
$\S$ and is invariant under $\diffsigma$. Given the transformation
(3.29a) for $X^a(\zeta)$, Eq.~(3.29b) is the required result for the
vanishing of the Poisson bracket $\{ P_a(\zeta) {X^a}_{,k}(\zeta) ,
H[\vec N] \}$.

The smeared momentum constraint (3.27) appears as a term in the action
functionals (3.1) and (3.17). The action (3.22)
employs the Lagrange multiplier $\boldN^k(\zeta)$ and defines a
smeared constraint
\be
H[\vec\boldN] := \int_\S d^3\zeta  \, \boldN^k(\zeta) \boldH_k(\zeta)
= \int_\Sigma d^3x \, \boldN^k(Z(x)) {X^a}_{,k}(Z(x)) H_a(x)
\eqnum{3.33} \ .
\ee
For any canonical variable $F$, the Poisson brackets
$\{ F,H[\vec\boldN]\}$ and $\{ F,H[\vec N]\}$ are
equal modulo the constraints $H_a(x) = 0 = \boldH_k(\zeta)$. Thus, the
smeared constraint (3.33) is the canonical generator of $\diffsigma$
on the constraint surface $H_a(x) = 0 = \boldH_k(\zeta)$.

\subsection{Noether Charges}

The Noether charges $Q[\bphi]$ and $Q[\btheta]$ from Section 2b can be
expressed in terms of the $\Sigma$--variables or the $\S$--variables:
\bea
Q[\bphi] & = &  \int_\Sigma d^3x \, \bphi(Z(x)) \Pi_\varphi(x)
= \int_\S d^3\zeta \, \bphi(\zeta)\bPi_\varphi(\zeta)\ ,\eqnum{3.34a} \\
Q[\btheta] & = &  \int_\Sigma d^3x \, \btheta(Z(x))\Pi_\vartheta(x)
= \int_\S d^3\zeta \,  \btheta(\zeta) \bPi_\vartheta(\zeta)
\ .\eqnum{3.34b}
\eea
$Q[\bphi]$ and $Q[\btheta]$ generate canonical transformations that yield
changes in the dynamical variables due to deformations of the hypersurfaces
of constant $\varphi$ and $\vartheta$, respectively. Among the
canonical $\S$--variables, only $\bvarphi$ is affected by $Q[\bphi]$ and
only $\bvartheta$ is affected by $Q[\btheta]$.

The Noether charge
\be
Q[\vec\bxi] = -\int_\Sigma d^3x\, \bxi^k(Z(x)) P_k(x) \eqnum{3.35}
\ee
from Eq.~(2.18) can be written in terms of the
$\S$--variables as follows. First, write the momentum
constraint as
\be
H_a(x)  = P_k(x) {Z^k}_{,a}(x) + H_a^+(x)
\ ,\eqnum{3.36}
\ee
where $H_a^+(x)$ contains the contributions from the
$\Sigma$ tensor fields $\varphi$, $\Pi_\varphi$, $\vartheta$,
$\Pi_\vartheta$, $g_{ab}$, and $p^{ab}$. This expression can be
solved for $P_k(x)$,
\be
P_k(X(\zeta)) = H_a(X(\zeta)) {X^a}_{,k}(\zeta) - H_a^+(X(\zeta))
{X^a}_{,k}(\zeta) \ ,\eqnum{3.37}
\ee
and the result inserted into Eq.~(3.35) along with a change of
integration variables $x^a = X^a(\zeta)$. The first term can be
rewritten using Eqs.~(3.15) and (3.23), while the second term
involves a straightforward mapping of $\Sigma$--tensor fields
to the fluid space. The result,
\be
Q[\vec\bxi] = \int_\S d^3\zeta \, \bxi^k(\zeta) \Bigl( P_a(\zeta)
{X^a}_{,k}(\zeta)  + \boldH_k^+(\zeta) \Bigr) \ ,\eqnum{3.38}
\ee
can be recognized as the canonical generator of fluid space
diffeomorphisms, $\diffs$. That is, for any canonical variable $F$,
the Poisson bracket $\{ F , Q[\vec\bxi]\}$ gives the change in $F$
due to an infinitesimal diffeomorphism of $\S$ generated by the
vector field $\vec\bxi(\zeta)$. If $F$ is one of the fluid space
tensor fields then the Poisson bracket $\{ F , Q[\vec\bxi]\}$
equals the Lie derivative $\pounds_{\vec\xi} F$. For the variables
$X^a(\zeta)$ and $P_a(\zeta)$, the transformation generated by
$Q[\vec\bxi]$ yields
\bea
\{ X^a(\zeta) , Q[\vec\bxi] \} & = & \bxi^k(\zeta) {X^a}_{,k}(\zeta)
\ ,\eqnum{3.39a}\\
\{ P_a(\zeta) , Q[\vec\bxi] \} & = & \Bigl( \bxi^k(\zeta) P_a(\zeta)
\Bigr)_{,k} \ .\eqnum{3.39b}
\eea
Thus, $X^a(\zeta)$ transforms as a set of scalar fields on $\S$ and
$P_a(\zeta)$ transforms as a set of scalar densities on $\S$.
\section{Perfect Fluid Action $\Sbar$}
\subsection{Routh's Procedure}

Given an equation of state $\rho(n,s)$ or $p(\mu,s)$ and the corresponding
definitions (2.15), the Hamiltonian constraint $\boldH_\perp(\zeta)$ of
Eq.~(3.18) depends explicitly on the canonical $\S$--variables
$\bg_{k\ell}(\zeta)$, $\bpi^{k\ell}(\zeta)$, and $\bPi_\varphi(\zeta)$,
and also on $s$, $\boldH_k^{\sss F}$, and $n$ or $\mu$ (depending on the
equation of state used). The variable $n$ or $\mu$ is determined
as a function of $\boldH_k^{\sss F}$, $s$, and the canonical
$\S$--variables $\bg_{k\ell}(\zeta)$ and $\bPi_\varphi(\zeta)$
through the auxiliary equation (3.19). The entropy per particle $s$ is
defined by $s := \bPi_\vartheta(\zeta)/\bPi_\varphi(\zeta)$,
and $\boldH_k^{\sss F}$ is determined as a function of the canonical
$\S$--variables $X^a(\zeta)$, $P_a(\zeta)$, $\bg_{k\ell}(\zeta)$, and
$\bpi^{k\ell}(\zeta)$ according to Eq.~(3.20). Thus,
the Hamiltonian constraint $\boldH_\perp(\zeta)$ depends on
the canonical $\S$--variables $\bg_{k\ell}(\zeta)$,
$\bpi^{k\ell}(\zeta)$, $X^a(\zeta)$, $P_a(\zeta)$,
$\bPi_\varphi(\zeta)$, and $\bPi_\vartheta(\zeta)$. It does
not depend on $\bvarphi(\zeta)$ or $\bvartheta(\zeta)$. Likewise,
the momentum constraint $\boldH_k(\zeta)$ of Eq.~(3.23) does not
depend on $\bvarphi(\zeta)$ or $\bvartheta(\zeta)$. It follows that
the Hamiltonian $\int_\S d^3\zeta \Bigl( \boldN^\perp(\zeta)
\boldH_\perp(\zeta)  + \boldN^k(\zeta) \boldH_k(\zeta)
\Bigr)$ for the combined perfect fluid/gravity system
does not depend on $\bvarphi(\zeta)$ or $\bvartheta(\zeta)$;
these variables are therefore ignorable (cyclic).

Since $\bvarphi(\zeta)$ and $\bvartheta(\zeta)$ are ignorable,
they and their canonical conjugates can be eliminated as dynamical
variables through the standard procedure due to Routh \cite{Lanczos}.
Accordingly, the equations of motion for $\bvarphi(\zeta)$
and $\bvartheta(\zeta)$ imply that $\bPi_\varphi(\zeta)$ and
$\bPi_\vartheta(\zeta)$ are constant in time. Then
$\bPi_\varphi(\zeta)$ and $\bPi_\vartheta(\zeta)$ can be interpreted
as fixed, prescribed fluid space densities, and the total time
derivative terms $\bPi_\varphi {\dot\varphi}$ and
$\bPi_\vartheta {\dot\vartheta}$ can be discarded from the action
(3.22). The resulting action is
\bea
\lefteqn{ {\bar S}[X^a,P_a,\bg_{k\ell},\bp^{k\ell},\boldN^\perp,
\boldN^k] } \qquad\quad\nonumber\\
& & = \int_{\rr}dt \int_\S d^3\zeta \biggl\{ P_a(\zeta){\dot X}^a(\zeta)
+ \bp^{k\ell}(\zeta) {\dot\bg}_{k\ell}(\zeta)
- \boldN^\perp(\zeta) \boldH_\perp(\zeta)  - \boldN^k(\zeta)
\boldH_k(\zeta) \biggr\} \ .\eqnum{4.1}
\eea
The Hamiltonian and momentum constraints that appear here are defined
by the same expressions (3.18), (3.23) as before, but now
$\bPi_\varphi(\zeta)$ and $\bPi_\vartheta(\zeta)$ are viewed as
fixed densities on the fluid space $\S$.

The symmetries (2.12b) and (2.12c) of the $S$ action are absent from
the new action $\Sbar$, and the corresponding charges (3.34) play no
role. On the other hand, the action $\Sbar$, like $S$, is invariant
under arbitrary diffeomorphisms of the fluid space $\S$ as long as
$\bPi_\varphi(\zeta)$ and $\bPi_\vartheta(\zeta)$  are transformed
as densities. In this sense the description of perfect fluids
provided by $\Sbar$ is independent of the choice of coordinates
$\zeta^k$ on $\S$. However, this invariance involves a
transformation of the fixed quantities $\bPi_\varphi(\zeta)$ and
$\bPi_\vartheta(\zeta)$ in addition to a transformation of the
dynamical variables. Therefore it does not, in general, correspond
to a conserved Noether charge. Only the subset of fluid space
diffeomorphisms $\vec\bxi$ that satisfy
\be
\pounds_{\vec\xi} \bPi_\varphi = 0 \ ,\qquad
\pounds_{\vec\xi} \bPi_\vartheta = 0  \eqnum{4.2}
\ee
give rise to a conserved charge \cite{Brown}. This charge is
given by expression (3.38) with the terms proportional
to $\bPi_\varphi(\zeta)$ and $\bPi_\vartheta(\zeta)$ dropped:
\be
Q[\vec\bxi] = \int_\S d^3\zeta \, \bxi^k(\zeta) \Bigl( P_a(\zeta)
{X^a}_{,k}(\zeta)  + \boldH_k^{\sss G}(\zeta) \Bigr) \ .\eqnum{4.3}
\ee
This charge generates a symmetry of $\Sbar$ through the Poisson
brackets for any fluid space vector $\vec\bxi$ that leaves
the particle number density $\bPi_\varphi(\zeta)$ and entropy
density $\bPi_\vartheta(\zeta)$ invariant
under Lie transport, Eq.~(4.2).
\subsection{$\Sigma$--Variables for $\Sbar$}

Now apply the point canonical transformation
\be
\Bigl( X^a(\zeta),\  \bg_{k\ell}(\zeta) \Bigr) \mapsto \Bigl( Z^k(x),\
g_{ab}(x) \Bigr)  \eqnum{4.4}
\ee
to the action (4.1). For the coordinates $X^a(\zeta)$,
$\bg_{k\ell}(\zeta)$, this is the inverse of the point transformation
(3.7) from the canonical $\Sigma$--variables to the canonical
$\S$--variables. The term $P_a(\zeta){\dot X}^a(\zeta) $ in the action
can be rewritten using Eq.~(3.10). Application of the relationship
(3.12) (derived in Appendix B) to the gravitational kinetic term leads to
\be
\int_\S d^3\zeta\, \bp^{k\ell}(\zeta) {\dot\bg}_{k\ell}(\zeta)
= \int_\Sigma d^3x\Bigl( p^{ab}(x) {\dot g}_{ab}(x) - H^{\sss G}_a(x)
{X^a}_{,k}(Z(x)) {\dot Z}^k(x) \Bigr) \ .\eqnum{4.5}
\ee
Together these results yield
\be
\int_\S d^3\zeta \Bigl( P_a(\zeta){\dot X}^a(\zeta)
+ \bp^{k\ell}(\zeta) {\dot\bg}_{k\ell}(\zeta) \Bigr) =
\int_\Sigma d^3x \Bigl( P_k(x) {\dot Z}^k(x)
+ p^{ab}(x) {\dot g}_{ab}(x) \Bigr) \ ,\eqnum{4.6}
\ee
where the momentum conjugate to $Z^k(x)$ is
\be
P_k(x) := - \biggl|\frac{\partial Z(x)}{\partial x}\biggr|
\Bigl( P_a(Z(x)) {X^a}_{,k}(Z(x)) + \boldH_k^{\sss G}(Z(x))
\Bigr) \ .\eqnum{4.7}
\ee
Note that the momentum $P_k(x)$ defined here differs from
the momentum (which was given the same notation $P_k(x)$) of
Section 2b.
This is clear from the simple observation that Eq.~(3.16),
which defines the relationship between $P_a(\zeta)$ and
the ``old" $P_k(x)$, contains the variables $\varphi$,
$\vartheta$, and their conjugates. These variables do not
appear in the relationship (4.7) between $P_a(\zeta)$ and
the ``new" $P_k(x)$. Therefore the new canonical
variables $Z^k(x)$, $P_k(x)$, $g_{ab}(x)$, and $p^{ab}(x)$
are not simply a subset of the canonical $\Sigma$--variables
that appear in the Hamiltonian version (3.1) of the
velocity--potential action $S$. Nevertheless, the set
$Z^k(x)$, $P_k(x)$, $g_{ab}(x)$, and $p^{ab}(x)$ defined
here will be referred to as $\Sigma$--variables in order to
distinguish them from the $\S$--variables $X^a(\zeta)$,
$P_a(\zeta)$, $\bg_{k\ell}(\zeta)$, and $\bp^{k\ell}(\zeta)$.

The canonical transformation (4.4) can be accompanied by a
change of variables (3.21) in which the Lagrange multipliers
$\boldN^\perp(\zeta)$ and $\boldN^k(\zeta)$ are replaced by
$N^\perp(x)$ and  $N^a(x)$. Then the action (4.1) becomes
\bea
\lefteqn{ {\bar S}[Z^k,P_k,g_{ab},p^{ab},N^\perp,
N^k] } \qquad\quad\nonumber\\
& & = \int_{\rr}dt \int_\Sigma d^3x \biggl\{ P_k(x){\dot Z}^k(x)
+ p^{ab}(x) {\dot g}_{ab}(x)
- N^\perp(x) H_\perp(x)  -  N^a(x) H_a(x) \biggr\} \ ,\eqnum{4.8}
\eea
where the momentum constraint is
\be
H_a(x) := \biggl|\frac{\partial Z(x)}{\partial x}\biggr|
\boldH_k(Z(x)) {Z^k}_{,a}(x) = P_k(x) {Z^k}_{,a}(x) + H_a^{\sss G}(x)
\ .\eqnum{4.9}
\ee
The perfect fluid contribution
\be
H_a^{\sss F}(x) := \biggl|\frac{\partial Z(x)}{\partial x}\biggr|
\boldH_k^{\sss F}(Z(x)) {Z^k}_{,a}(x) = P_k(x) {Z^k}_{,a}(x) \ ,
\eqnum{4.10}
\ee
differs from the previous expression (2.14a) in that the terms
involving the coordinates $\varphi$, $\vartheta$, and their
conjugates are absent.

The Hamiltonian constraint in the action (4.8) can be written in
terms of the $\Sigma$--variables as
\be
H_\perp(x) := \biggl|\frac{\partial Z(x)}{\partial x}\biggr|
\boldH_\perp(Z(x)) = H_\perp^{\sss G}(x) + \sqrt{\mu^2 \Pi_\varphi^2 +
H_a^{\sss F} g^{ab} H_b^{\sss F} } - \sqrt{g} p \ .\eqnum{4.11}
\ee
Here, $\Pi_\varphi(x)$ is a function of $Z(x)$ defined by
\be
\Pi_\varphi(x) := \biggl|\frac{\partial Z(x)}{\partial x}\biggr|
\bPi_\varphi(Z(x)) \ ,\eqnum{4.12}
\ee
with $\bPi_\varphi(\zeta)$ a fixed density on $\S$.
In $H_\perp(x)$, $\mu$ and $p$ are determined as functions
of $s$ and either $n$ or $\mu$ by the equation of state, either
$\rho(n,s)$ or $p(\mu,s)$, and the associated
definitions (2.15). The entropy per particle $s$ is a
function of $Z(x)$, $s = s(Z(x))$, as determined by the fixed
scalar function
\be
s(\zeta) := \frac{\bPi_\vartheta(\zeta)}{\bPi_\varphi(\zeta)}
\eqnum{4.13}
\ee
on the fluid space $\S$. The auxiliary equation
\be
\frac{\Pi_\varphi}{\sqrt{g}} = n \sqrt{ 1 + \frac{H^{\sss F}_a g^{ab}
H^{\sss F}_b}{ \mu^2 \Pi_\varphi^2} }  \eqnum{4.14}
\ee
determines either $n$ or $\mu$, depending on the equation of
state.
The auxiliary equation here is formally identical to Eq.~(2.16),
but now $H_a^{\sss F}(x)$ and $\Pi_\varphi(x)$ are defined by
Eqs.~(4.10) and (4.12), respectively. Thus, $n$ or $\mu$ is determined
as a function of $g_{ab}(x)$, $P_k(x)$ (which appears in
$H_a^{\sss F}(x)$), and $Z^k(x)$ (which appears in $\Pi_\varphi(x)$,
$s(Z(x))$, and $H_a^{\sss F}(x)$).

With the help of Eq.~(4.7) the conserved charge (4.3) can be written
in terms of the $\Sigma$--variables as
\be
Q[\vec\bxi] = -\int_\Sigma d^3x \, \bxi^k(Z(x)) P_k(x) \ .\eqnum{4.15}
\ee
Again, $\vec\bxi$ must satisfy the conditions (4.2). The symmetry
generated by $Q[\vec\bxi]$ with such a vector $\vec\bxi$ corresponds
to the transformations $\delta Z^k = - \bxi^k(Z)$ and
$\delta P_k = P_\ell {\bxi^\ell}_{,k}(Z)$.
\subsection{Lagrangian $\Sbar$ Action}

The Lagrangian form of the action (4.8) is obtained by eliminating
the momenta $P_k(x)$ and $p^{ab}(x)$. For the gravitational variables,
elimination of $p^{ab}(x)$ yields the usual Einstein--Hilbert
action. For the perfect fluid variables, variation of the action
with respect to $P_k(x)$ gives the equation of motion
\be
0 = {\dot Z}^k - N^a{Z^k}_{,a} - N^\perp\sqrt{g} \left(
n\frac{\partial\mu}{\partial P_k} - \frac{\partial p}{\partial P_k}
+ \frac{n H^{\sss F}_a g^{ab} {Z^k}_{,b}}{\mu \Pi_\varphi^2}
\right) \ ,\eqnum{4.16}
\ee
where the auxiliary equation (4.14) has been used to
eliminate the expression $H_a^{\sss F} g^{ab} H_b^{\sss F}$. By
applying the relationships (2.15) and using the fact that $s=s(Z(x))$
is independent of $P_k(x)$, one finds that the two terms involving
$\partial\mu/\partial P_k$ and $\partial p/\partial P_k$ cancel one
another. Thus, the $P_k(x)$ equation of motion becomes
\be
{\dot Z}^k - N^a{Z^k}_{,a} = N^\perp\sqrt{g} n
\frac{H^{\sss F}_a g^{ab} {Z^k}_{,b}}{\mu \Pi_\varphi^2}
\ .\eqnum{4.17}
\ee
With this result, the perfect fluid contribution to the action (4.8)
reduces to
\be
\Sbar^{\sss F}[Z^k;g_{ab},N^\perp,N^a] = \int_{\rr} dt
\int_\Sigma d^3x\, N^\perp\sqrt{g} (p - n\mu) \  .\eqnum{4.18}
\ee
Here, $n$, $\mu$, and $p$ are determined by the equation of state,
the relationships (2.15), and the solution of the auxiliary
equation (4.14). In the auxiliary equation, $P_k(x)$ is replaced
by the solution of the equation of motion (4.17).

The action (4.18) can be put in covariant form by interpreting
$t$ and $x^a$ as spacetime coordinates $y^0=t$, $y^a = x^a$.
The spacetime metric $\gamma_{\alpha\beta}$ is constructed
from the spatial metric, lapse, and shift according to the
usual ADM decomposition,
\be
\gamma_{\alpha\beta} dy^\alpha dy^\beta = -(N^\perp)^2 dt^2
+ g_{ab} (N^a dt + dx^a)(N^b dt + dx^b) \ .\eqnum{4.19}
\ee
This implies $N^\perp\sqrt{g} = \sqrt{-\gamma}$.

If the equation of state is specified by the function
$\rho(n,s)$, the relationships (2.15) show that
the covariant action $\Sbar^{\sss F}$ becomes
\be
\Sbar^{\sss F}[Z^k;\gamma_{\alpha\beta}] = -\int d^4y
\sqrt{-\gamma} \rho(n,s) \ .\eqnum{4.20}
\ee
Here, the entropy per particle is a given function of the fluid
Lagrangian coordinates, $s=s(Z(y))$. In order to provide for a
covariant specification of $n$, first rewrite Eq.~(4.12) as
\bea
\Pi_\varphi & = & \sqrt{g} \epsilon^{abc}
{Z^1}_{,a}{Z^2}_{,b}{Z^3}_{,c} \bPi_\varphi(Z) \nonumber\\
& = & \frac{1}{3!} \sqrt{g} \epsilon^{abc} \eta_{k\ell m}(Z)
{Z^k}_{,a}{Z^\ell}_{,b}{Z^m}_{,c} \ ,\eqnum{4.21}
\eea
where $\epsilon^{abc}$ is obtained by raising indices on
the volume form on $\Sigma$.
Also, $\eta_{k\ell m}$ are the components of a three--form
\be
\eta = \frac{1}{3!} \eta_{k\ell m} dZ^k\wedge dZ^\ell\wedge
dZ^m = \bPi_\varphi d^3Z  \eqnum{4.22}
\ee
on the fluid space $\S$, so that $\eta_{123}(Z) = \bPi_\varphi(Z)$
is the fixed density on $\S$. Now consider the spacetime vector
density
\be
J^\alpha := - \frac{1}{3!}\sqrt{-\gamma}
\epsilon^{\alpha\beta\sigma\rho} \eta_{k\ell m}
{Z^k}_{,\beta}{Z^\ell}_{,\sigma}{Z^m}_{,\rho} \ ,\eqnum{4.23}
\ee
where $\epsilon^{\alpha\beta\sigma\rho}$ is obtained by
raising indices on the spacetime volume form. Note that $J^\alpha$
is tangent to the fluid flow lines, since it is orthogonal to
the gradients of the Lagrangian coordinates. The $t$ component
of $J^\alpha$ is $J^t = \Pi_\varphi$, while the spatial components
are
\be
J^a = -\frac{1}{2} \sqrt{g} \epsilon^{abc}\eta_{k\ell m} {\dot Z}^k
{Z^\ell}_{,b} {Z^m}_{,c} \ .\eqnum{4.24}
\ee
Now, the equation of motion (4.17) for $P_k(x)$ can be used to
obtain the result
\be
J^a + N^a J^t = -\left( \frac{N^\perp \sqrt{g}n}{\mu}\right)
\frac{g^{ab}H_b^{\sss F}}{J^t} \ .\eqnum{4.25}
\ee
Using the ADM split (4.19), the squared magnitude of $J^\alpha$
becomes
\bea
|J|^2 := -J^\alpha \gamma_{\alpha\beta} J^\beta & = &
(N^\perp J^t)^2  - g_{ab}(J^a + N^aJ^t)(J^b + N^bJ^t)
\nonumber\\ & = & (n\sqrt{-\gamma})^2  \eqnum{4.26}
\eea
where, again, the auxiliary equation (4.14) has been used.
Thus, $J^\alpha$ is the densitized particle number flux vector,
$J^\alpha = n \sqrt{-\gamma} U^\alpha$.
In the action (4.20), $n$ is the function of $Z^k$ defined by
$n := |J|/\sqrt{-\gamma}$,
with $J^\alpha$ defined by Eq.~(4.23). The action (4.20) is
discussed in Ref.~\cite{Brown}, where it is
confirmed that the variation $\delta{\bar S}^{\sss F}$ leads
to the correct perfect fluid stress--energy--momentum tensor
and equations of motion.

The Lagrangian action $\Sbar$, like its canonical counterpart,
is invariant under fluid space diffeomorphisms that leave
the particle number density and entropy density invariant.
That is, the action $\Sbar$ is invariant under transformations
$\delta Z^k = - \bxi^k(Z)$ that satisfy
\be
\pounds_{\vec\bxi} \eta = 0 \ ,\qquad \pounds_{\vec\bxi} s = 0
\ .\eqnum{4.27}
\ee
Indeed, with the definitions of $n$ and $s$ as functions
of $Z^k$, it follows that \cite{Brown}
\bea
\delta n & = & - \frac{1}{3!} U_\alpha
\epsilon^{\alpha\beta\sigma\rho} (\pounds_{\vec\bxi}
\eta_{k\ell m}) {Z^k}_{,\beta}{Z^\ell}_{,\sigma}
{Z^m}_{,\rho} \ ,\eqnum{4.28a} \\
\delta s & = & - \pounds_{\vec\bxi} s \ .\eqnum{4.28b}
\eea
Thus, $\delta n$ and $\delta s$, and in turn $\delta\Sbar$,
vanish for $\vec\bxi$ satisfying Eq.~(4.27).

If the equation of state is specified by the function $p(\mu,s)$,
then the integrand in the action (4.18) can be expressed
as the product of $\sqrt{-\gamma}$ and $p(\mu,s) - \mu\,n(\mu,s)$,
where $n(\mu,s) := \partial p(\mu,s)/\partial\mu$. The chemical
potential $\mu$ can be defined by the solution of the equation
$n(\mu,s) = |J|/\sqrt{-\gamma}$. This gives $\mu(n,s)$, where $n$
is now viewed as shorthand notation for $|J|/\sqrt{-\gamma}$.
By substituting  $\mu(n,s)$ for $\mu$, one finds that the integrand
reduces to the product of $\sqrt{-\gamma}$ and  $p(\mu(n,s),s)
- \mu(n,s)\, n(\mu(n,s),s) = - \rho(n,s)$, and the resulting action
simply coincides with the action (4.20) described above.
Note that the process of solving
for $\mu(n,s)$ and substituting the result back into the
integrand is equivalent to performing a Legendre transformation
in which the equation of state is changed from $p(\mu,s)$ to
$\rho(n,s)$. Thus, we reach the obvious conclusion that
a Lagrangian $\Sbar$--type action for a perfect fluid
with a given equation of state $p(\mu,s)$ can be obtained by
rewriting the equation of state as a function $\rho(n,s)$ and
then using the action (4.20).
\subsection{Hybrid Action
$\Sbar^{\sss F}[Z^k,\vartheta;\gamma_{\alpha\beta}]$}

It is possible, of course, to eliminate just one of the ignorable
coordinates, either $\bvarphi$ or $\bvartheta$, from the action $S$
while retaining the other as a dynamical variable. This leads to a
hybrid action that is, loosely speaking, half way between the
velocity--potential type action $S$ and the action $\Sbar$ discussed
above. In this subsection I will treat the case in which $\bvarphi$
and its conjugate are eliminated, and $\bvartheta$ and its conjugate
are retained. This leads to an action of the $\Sbar$ type in the
sense that the fluid four--velocity is expressed in terms of the
totally antisymmetric product of gradients of $Z^k$, as in Eq.~(1.3).

The variables $\bvarphi(\zeta)$ and $\bPi_\varphi(\zeta)$ are eliminated
from the action (3.22) by dropping the term
$\bPi_\varphi {\dot\bvarphi}$ and elsewhere interpreting
$\bPi_\varphi(\zeta)$ as a fixed density on the fluid space $\S$.
The resulting action can be expressed in terms of $\Sigma$--variables
by applying the point canonical transformation
\be
\Bigl( X^a(\zeta),\  \bvartheta(\zeta),\  \bg_{k\ell}(\zeta) \Bigr)
\mapsto \Bigl( Z^k(x),\  \vartheta(x),\  g_{ab}(x) \Bigr)
\ .\eqnum{4.29}
\ee
The momenta conjugate to $\vartheta(x)$, $g_{ab}(x)$, and $Z^k(x)$
are $\Pi_\vartheta(x)$, $p^{ab}(x)$, and
\be
P_k(x) := - \left|\frac{\partial Z(x)}{\partial x}\right|  \left(
P_a(Z(x)) {X^a}_{,k}(Z(x)) + \bPi_\vartheta(Z(x))\bvartheta_{,k}(Z(x))
+ \boldH_k^{\sss G}(Z(x)) \right) \ ,\eqnum{4.30}
\ee
respectively. Associated with the
Lagrange multiplier $N^a(x)$ is the momentum
constraint $H_a(x) = H_a^{\sss F}(x) + H_a^{\sss G}(x)$ with
the perfect fluid contribution
\be
H_a^{\sss F}(x) = P_k(x) {Z^k}_{,a}(x) + \Pi_\vartheta(x)
\vartheta_{,a}(x) \ .\eqnum{4.31}
\ee
The Hamiltonian constraint associated with the Lagrange multiplier
$N^\perp(x)$ is given by Eqs.~(4.11), (4.12), and
(4.14). In this case the entropy per particle,
\be
s := \frac{\Pi_\vartheta(x)}{\Pi_\varphi(x)} \ ,\eqnum{4.32}
\ee
is a function of $\Pi_\vartheta(x)$ and $Z^k(x)$ (through
$\Pi_\varphi(x)$).

The symmetries of this action are generated through the Poisson
brackets by the conserved charges
\bea
Q[\btheta] & = & \int_\Sigma d^3x \, \Pi_\vartheta(x) \btheta(Z(x))
\ ,\eqnum{4.33a} \\
Q[\vec\bxi] & = & \int_\Sigma d^3x \, P_k(x)  \bxi^k(Z(x))
\ ,\eqnum{4.33b}
\eea
where $\vec\bxi$ leaves the fluid space number density invariant,
$\pounds_{\vec\bxi}\bPi_\varphi = 0$. The charge
(4.33a) generates a deformation of the constant $\vartheta$
surfaces, $\delta\vartheta = \btheta(Z)$, while the charge
(4.33b) generates a fluid space diffeomorphism,
$\delta Z^k = -\bxi^k(Z)$.

The action can be written in Lagrangian form by eliminating
the momenta $P_k(x)$, $\Pi_\vartheta(x)$, and $p^{ab}(x)$.
Elimination
of $p^{ab}(x)$ leads to the usual Einstein--Hilbert action for the
gravitational field. Varying the action with respect to
$P_k(x)$ produces the equation of motion (4.16), which reduces to
Eq.~(4.17). With this result, the perfect fluid contribution to the
action becomes
\bea
\lefteqn{ \Sbar^{\sss F}[Z^k,\vartheta,\Pi_\vartheta;g_{ab},
N^\perp,N^a] } \qquad\quad\nonumber\\
& & = \int_{\rr}dt \int_\Sigma d^3x \biggl\{ N^\perp\sqrt{g}
(p - n\mu) + \Pi_\vartheta \biggl( {\dot\vartheta} -
N^a\vartheta_{,a} - N^\perp\sqrt{g} \frac{n H_a^{\sss F} g^{ab}
\vartheta_{,b}}{\mu \Pi_\varphi^2} \biggr) \biggr\} \ .\eqnum{4.34}
\eea
At this stage it is convenient to make a change of variables in which
$s = \Pi_\vartheta/\Pi_\varphi$ replaces $\Pi_\vartheta$ as an
independent variable. (Thus, for the moment, the momentum
$\Pi_\vartheta$ will not be eliminated from the action, but
will be hidden in the new variable $s$.) Assuming now an equation
of state of the form $\rho(n,s)$, the action is
\bea
\lefteqn{ \Sbar^{\sss F}[Z^k,\vartheta,s;g_{ab},
N^\perp,N^a] } \qquad\qquad\nonumber\\
& & = \int d^4y \biggl\{ -N^\perp\sqrt{g}
\rho(n,s) + s\Pi_\varphi \biggl( {\dot\vartheta} -
N^a\vartheta_{,a} - N^\perp\sqrt{g} \frac{n H_a^{\sss F} g^{ab}
\vartheta_{,b}}{\mu \Pi_\varphi^2} \biggr) \biggr\} \ .\eqnum{4.35}
\eea
Again, define the vector density $J^\alpha$ of Eq.~(4.23).
With the result (4.25) it is straightforward to verify that
\be
J^\alpha \vartheta_{,\alpha} = \Pi_\varphi\left( {\dot\vartheta}
- N^a \vartheta_{,a} - N^\perp\sqrt{g} \frac{n H_a^{\sss F} g^{ab}
\vartheta_{,b}}{\mu \Pi_\varphi^2} \right) \ ,\eqnum{4.36}
\ee
so that the action simplifies to
\be
\Sbar^{\sss F}[Z^k,\vartheta,s;\gamma_{\alpha\beta}] =
\int d^4y \Bigl( -\sqrt{-\gamma} \rho(n,s) +
sJ^\alpha\vartheta_{,\alpha} \Bigr) \ .\eqnum{4.37}
\ee
Here, $n = |J|/{\sqrt{-\gamma}}$
depends on $Z^k$ through the definition (4.23) for $J^\alpha$.

The entropy per particle $s$ (and hence the momentum $\Pi_\vartheta$)
can be eliminated from the action (4.37) by use of the solution
of the $s$ equation of motion, namely,
\be
0 = -\sqrt{-\gamma} \frac{\partial\rho(n,s)}{\partial s} +
J^\alpha \vartheta_{,\alpha} \ .\eqnum{4.38}
\ee
{}From the first law Eq.~(2.1), we have $\partial\rho/\partial s =
n T$ so that $\vartheta$ is related to temperature
by $T = \vartheta_{,\alpha} U^\alpha$ (Eq.~(2.5)).  It follows
that the solution of the $s$ equation of motion (4.38) has
the form $s(n,T)$, the entropy per particle as a function of
number density $n$ and temperature $T$. If this result is
substituted into the action (4.37), the integrand becomes the
product of $\sqrt{-\gamma}$ and a function of $n$ and $T$,
\be
n\, a(n,T) := nT\,s(n,T) - \rho(n,s(n,T)) \ .\eqnum{4.39}
\ee
The function $a(n,T)$ is the physical free energy of the fluid.
The action (4.37) therefore reduces to
\be
\Sbar^{\sss F}[Z^k,\vartheta;\gamma_{\alpha\beta}] =
-\int d^4y \sqrt{-\gamma}\, n\,a(n,T) \ ,\eqnum{4.40}
\ee
where $n=|J|/{\sqrt{-\gamma}}$
and $T = \vartheta_{,\alpha} J^\alpha/|J|$
with $J^\alpha$ defined as a function of $Z^k$ by Eq.~(4.23).
This action is invariant under the symmetry transformations
$\delta\vartheta = \btheta(Z)$ and $\delta Z^k = -\bxi^k(Z)$,
where $\vec\bxi$ leaves the fluid space three--form $\eta$
invariant under Lie transport, $\pounds_{\vec\bxi}\eta = 0$.

The action (4.40) is the one discussed by Kijowski
{\it et al.}~\cite{Kijowski}. The equation of state
is specified by giving the physical free energy $a$ as a function
of number density $n$ and temperature $T$.
\appendix
\section{Hamiltonian Action with Equation of State $p(\mu,\lowercase{s})$}

Let $t$ denote a scalar function on spacetime whose gradient is nonzero
and (future pointing) timelike. The $t={\rm constant}$ surfaces foliate
spacetime into spacelike hypersurfaces with unit normal $n_\alpha =
-N^\perp t_{,\alpha}$, where $N^\perp := (-t_{,\alpha} \gamma^{\alpha\beta}
t_{,\beta})^{-1/2}$ defines the lapse function. Also introduce the time
vector field $t^\alpha$ such that $t_{,\alpha} t^\alpha = 1$ and the
projection tensor $g^\alpha_\beta := \delta^\alpha_\beta + n^\alpha n_\beta$
onto the leaves of the foliation. With the shift vector defined by
$N^\alpha := g^\alpha_\beta t^\beta$, the time vector field can be
written as $t^\alpha = N^\perp n^\alpha + N^\alpha$.

Using the above relationships, the vector $V^\alpha$ can be decomposed as
\be
V^\alpha = (g^\alpha_\beta - n^\alpha n_\beta)V^\beta
= (gV)^\alpha + ( t^\alpha - N^\alpha) V^t \ ,\eqnum{A.1}
\ee
where the definitions $(gV)^\alpha := g^\alpha_\beta V^\beta$
and $V^t := t_{,\alpha} V^\alpha$ are used. It follows that the
chemical potential $|V|$ can be written as
\be
|V| := \sqrt{-V^\alpha \gamma_{\alpha\beta} V^\beta} =
\sqrt{(N^\perp V^t)^2 - (gV)_\alpha (gV)^\alpha} \ .\eqnum{A.2}
\ee
The action (2.9) then becomes
\be
S^{\sss F} = \int d^4y \sqrt{-\gamma} \biggl\{ -\rho(|V|,s) +
\frac{n(|V|,s)}{|V|} \bigl( V^t t^\alpha + (gV)^\alpha -
V^t N^\alpha\bigr) \bigl(\varphi_{,\alpha} + s\vartheta_{,\alpha}
- W_k {Z^k}_{,\alpha}\bigr) \biggr\} \ ,\eqnum{A.3}
\ee
where the definitions $n(\mu,s) := \partial p(\mu,s)/\partial\mu$ and
$\rho(\mu,s) := \mu n(\mu,s) - p(\mu,s)$ have been used.

Now tie the spacetime coordinates to the foliation by choosing $t$ as the
time coordinate and choosing spatial coordinates such that
$\partial/\partial t$ is the time vector field $t^\alpha$. The
$t$--components of $(gV)^\alpha$ and $N^\alpha$ vanish, and the
action becomes
\bea
\lefteqn{ S^{\sss F}[(gV)^a,V^t,\varphi,\vartheta,s,W_k,Z^k;
N^\perp,N^a,g_{ab}] }\qquad\quad\nonumber\\
& = & \int_{\rr}dt \int_\Sigma d^3x\, N^\perp\sqrt{g} \biggl\{
- \rho(|V|,s) + \frac{n(|V|,s)}{|V|} V^t ( {\dot\varphi}
+ s{\dot\vartheta} - W_k{\dot{Z}}^k) \nonumber\\
& & \qquad\qquad\qquad\qquad\quad + \frac{n(|V|,s)}{|V|} \bigl(
(gV)^a - V^tN^a\bigr) ( \varphi_{,a} + s\vartheta_{,a} - W_k {Z^k}_{,a} )
\biggr\} \ .\eqnum{A.4}
\eea
The form of this expression suggests a change of variables in which
$V^t$, $s$, and $W_k$ are replaced by
\bea
\Pi_\varphi & := & N^\perp\sqrt{g} \frac{n(|V|,s)}{|V|} V^t \ ,\eqnum{A.5a}\\
\Pi_\vartheta & := & N^\perp\sqrt{g} \frac{n(|V|,s)}{|V|} V^t s
\ ,\eqnum{A.5b}\\
P_k & := & -N^\perp\sqrt{g} \frac{n(|V|,s)}{|V|} V^t W_k\ .\eqnum{A.5c}
\eea
These variables will become the momenta conjugate to $\varphi$,
$\vartheta$, and $Z^k$, respectively.

Now introduce the variables
\be
\xi^a := \frac{(gV)^a}{V^t} \ ,\eqnum{A.6}
\ee
which can be viewed as replacements for $(gV)^a$. In
terms of the new variables (A.5), (A.6), the action becomes
\bea
\lefteqn{ S^{\sss F}[\xi^a,\varphi,\Pi_\varphi,\vartheta,\Pi_\vartheta,
Z^k,P_k;N^\perp,N^a,g_{ab}] }\qquad\nonumber\\
& = & \int_{\rr}dt \int_\Sigma d^3x \left\{ \Pi_\varphi {\dot\varphi}
+ \Pi_\vartheta {\dot\vartheta} + P_k{\dot Z}^k - N^\perp\sqrt{g}
\rho(|V|,\Pi_\vartheta/\Pi_\varphi)
+ (\xi^a - N^a) H^{\sss F}_a \right\} \ ,\eqnum{A.7}
\eea
where $s = \Pi_\vartheta/\Pi_\varphi$ (as follows from Eqs.~(A.5a) and
(A.5b))
and $H_a^{\sss F}$ is the fluid contribution to the momentum constraint,
\be
H_a^{\sss F} := \Pi_\varphi \varphi_{,a} + \Pi_\vartheta \vartheta_{,a}
+ P_k {Z^k}_{,a} \ .\eqnum{A.8}
\ee
In the action (A.7) $|V|$ is determined as follows. Definition (A.6) can
be combined with Eq.~(A.2) to yield
\be
V^t = \frac{|V|}{\sqrt{(N^\perp)^2 - \xi^a\xi_a}} \ .\eqnum{A.9}
\ee
Then the momentum (A.5a) becomes
\be
\Pi_\varphi = N^\perp \sqrt{g} \frac{n(|V|,\Pi_\vartheta/\Pi_\varphi)}
{\sqrt{(N^\perp)^2 - \xi^a \xi_a }} \ .\eqnum{A.10}
\ee
This can be viewed as an equation that determines $|V|$ as a function
of $\Pi_\varphi$, $\Pi_\vartheta$, $\xi^a$, $N^\perp$, and $g_{ab}$.

The next step is to eliminate $\xi^a$ from the action (A.7) by
substituting the solution of the $\xi^a$ equations of motion. The
$\xi^a$ equations of motion are found by varying $S$ with respect to
$\xi^a$, keeping the other variables fixed. For this calculation, the
variation of $|V|$ is needed. From Eq.~(A.10) this is determined to be
\be
\frac{\partial n(|V|,\Pi_\vartheta/\Pi_\varphi)}{\partial|V|} \delta|V|
= -\left(\frac{n(|V|,\Pi_\vartheta/\Pi_\varphi)}
{ (N^\perp)^2 - \xi^b \xi_b}\right) \xi_a\delta\xi^a \ .\eqnum{A.11}
\ee
Using this result and the relationship $\partial\rho(\mu,s)/\partial\mu
= \mu\partial n(\mu,s)/\partial\mu$ in the variation of the action (A.7),
the $\xi^a$ equations of motion are obtained:
\be
\frac{|V|}{\sqrt{(N^\perp)^2 - \xi^b\xi_b}} \xi_a =
- \frac{H_a^{\sss F}}{\Pi_\varphi} \ .\eqnum{A.12}
\ee
Here, $|V|$ is considered to be a function of $\Pi_\varphi$,
$\Pi_\vartheta$, $\xi^a$, $N^\perp$, and $g_{ab}$ through the solution
of Eq.~(A.10). Equation~(A.12) can be squared and, after a bit of
algebra, written as
\be
\frac{N^\perp}{\sqrt{(N^\perp)^2 - \xi^b\xi_b}} = \sqrt{1 +
\frac{H_a^{\sss F} g^{ab} H_b^{\sss F}}{|V|^2\Pi_\varphi^2}}
\ .\eqnum{A.13}
\ee
Then Eq.~(A.10) becomes
\be
\frac{\Pi_\varphi}{\sqrt{g}} = n(|V|,\Pi_\vartheta/\Pi_\varphi)
\sqrt{1 + \frac{H_a^{\sss F} g^{ab} H_b^{\sss F}}{|V|^2\Pi_\varphi^2}}
\ .\eqnum{A.14}
\ee
This equation determines $|V|$ as a function of $\Pi_\varphi$,
$\Pi_\vartheta$, $H^{\sss F}_a$, and $g_{ab}$.

The implicit solution (A.12) of the $\xi^a$ equations of motion
can be inserted into the action (A.7) to eliminate the variable
$\xi^a$. Note that $\xi^a$ only appears in the fluid contribution
to the Hamiltonian constraint; this is
\be
H^{\sss F}_\perp = \sqrt{g}|V|n(|V|,\Pi_\vartheta/\Pi_\varphi)
-\frac{\xi^a H_a^{\sss F}}{N^\perp} -
\sqrt{g} p(|V|,\Pi_\vartheta/\Pi_\varphi) \ ,\eqnum{A.15}
\ee
where $\rho(\mu,s) = \mu n(\mu,s) - p(\mu,s)$ has been used.
The term involving the number density $n$ can be rewritten
using Eq.~(A.10), and the term involving $\xi^a H_a^{\sss F}$
can be rewritten by contracting Eq.~(A.12) with $\xi^a$.
This leads to the expression
\be
H^{\sss F}_\perp = \Pi_\varphi |V|
\frac{N^\perp}{\sqrt{(N^\perp)^2 - \xi^b\xi_b}} - \sqrt{g}
p(|V|,\Pi_\vartheta/\Pi_\varphi) \ ,\eqnum{A.16}
\ee
which, when combined with Eq.~(A.13), becomes
\be
H^{\sss F}_\perp = \sqrt{|V|^2 \Pi_\varphi^2 +
H_a^{\sss F} g^{ab} H_b^{\sss F} } - \sqrt{g}
p(|V|,\Pi_\vartheta/\Pi_\varphi) \ .\eqnum{A.17}
\ee
Here, $|V|$ is determined as a function of $\Pi_\varphi$,
$\Pi_\vartheta$, $H^{\sss F}_a$, and $g_{ab}$ through
Eq.~(A.14).

With the result (A.17) the action (A.7) is brought to
standard Hamiltonian form with contributions (A.17) and
(A.8) to the Hamiltonian and momentum constraints. The
chemical potential $|V|$ that appears in $H_\perp^{\sss F}$
is determined by the auxiliary equation (A.14). This
agrees with the action described in
Eqs.~(2.13)--(2.17) (for the equation of state $p(\mu,s)$)
of the main text.
\section{Derivation of Equation (3.12)}

In this appendix the parameter dependence of tensors and mappings will
be denoted by subscripts. Thus, for example, the one--parameter family
of mappings from $\S$ to $\Sigma$ is denoted $X_t$, and the generic
one--parameter family of tensors on $\Sigma$ is denoted $\chi_t$.

Let $V_t$ denote the vector field on $\Sigma$ defined in terms of local
coordinates by
\be
V^a_t(x) :=  \biggl( \frac{d}{dt} X^a_t(\zeta) \biggr)
\biggr|_{\zeta=Z_t(x)}
\ .\eqnum{B.1}
\ee
For each value of $t$, the integral curves of $V_t$ define a
one--parameter $\sigma$ family of diffeomorphisms
${\cal N}_{t,\sigma}:\Sigma\to\Sigma$. By
definition ${\cal N}_{t,\sigma}$ satisfies the differential equation
\be
\frac{d{\cal N}_{t,\sigma}}{d\sigma} = V_t({\cal
N}_{t,\sigma}) \ ,\eqnum{B.2}
\ee
where ${\cal N}_{t,0} = I_\Sigma$ is the identity mapping on $\Sigma$.
By combining the previous two equations, we obtain
\be
\frac{d}{d\sigma} {\cal N}^a_{t,\sigma}(x) \biggr|_{\sigma=0}
= \biggl( \frac{d}{dt} X^a_t(\zeta) \biggr) \biggr|_{\zeta=Z_t(x)}
= \frac{d}{d\sigma} X^a_{t+\sigma}(Z_t(x)) \biggr|_{\sigma=0}
\ .\eqnum{B.3}
\ee
It readily follows that up to order $\sigma^2$, ${\cal N}_{t,\sigma}$
is given by
\be
{\cal N}_{t,\sigma} = X_{t+\sigma}\circ Z_t + {\cal O}(\sigma^2) \ .
\eqnum{B.4}
\ee
The ${\cal O}(\sigma^2)$ corrections to ${\cal N}_{t,\sigma}$ are
not needed in the analysis that follows.

Consider first the case in which $\chi_t$ is a one--parameter family of
scalar fields on $\Sigma$. The Lie derivative of $\chi_t$ with respect
to $V_t$ is given by
\be
\Bigl(\pounds_{V_t} \chi_t\Bigr)\biggr|_{x} :=
\frac{d}{d\sigma} \chi_t({\cal N}_{t,\sigma}(x)) \biggr|_{\sigma=0} =
\frac{d}{d\sigma} \chi_t(X_{t+\sigma}(\zeta)) \biggr|_{\sigma=0,
\ \zeta=Z_t(x)} \ .\eqnum{B.5}
\ee
In order to simplify this expression, denote the pullback of $\chi_t(x)$
to the fluid space $\S$ by $\bchi_t(\zeta) := \chi_t(X_t(\zeta))$ and
compute
\bea
\frac{d}{dt}\bchi_t(\zeta) & = & \frac{d}{d\sigma}
\bchi_{t+\sigma}(\zeta) \biggr|_{\sigma=0} \nonumber\\
& = & \frac{d}{d\sigma}
\chi_{t+\sigma}(X_{t+\sigma}(\zeta)) \biggr|_{\sigma=0} \nonumber\\
& = & \biggl( \frac{d}{dt} \chi_t(x) \biggr) \biggr|_{x=X_t(\zeta)}
+ \frac{d}{d\sigma}  \chi_t(X_{t+\sigma}(\zeta)) \biggr|_{\sigma=0}
\ .\eqnum{B.6}
\eea
Thus, the Lie derivative (B.5) equals
\be
\Bigl(\pounds_{V_t} \chi_t\Bigr)\biggr|_{x}  =
- \frac{d}{dt} \chi_t(x) +
\biggl(\frac{d}{dt} \bchi_t(\zeta) \biggr) \biggr|_{\zeta=Z_t(x)}
\ .\eqnum{B.7}
\ee
This result expresses the Lie derivative of $\chi_t$ along
$V_t$ in terms of the derivatives of $\chi_t$ and $\bchi_t$ with
respect to the parameter $t$.

The result (B.7) can be extended to arbitrary one--parameter families of
tensor fields $\chi_t$. For example, consider $\chi_t(x)$ to be a covariant
vector and compute
\bea \Bigl(\pounds_{V_t} \chi_t\Bigr)\biggr|_{x} & := &
\frac{d}{d\sigma} \biggl( {\cal N}^*_{t,\sigma}
\chi_t({\cal N}_{t,\sigma}(x)) \biggr)  \biggr|_{\sigma=0}  \nonumber\\
 & = & Z^*_t \frac{d}{d\sigma} \biggl( X^*_{t+\sigma}
 \chi_t(X_{t+\sigma}(\zeta))
 \biggr) \biggr|_{\sigma=0,\ \zeta=Z_t(x)}  \ .\eqnum{B.8}
\eea
Here, Eq.~(B.4) has been used to express
\be
{\cal N}^*_{t,\sigma} = Z_t^*\circ X_{t+\sigma}^* +
{\cal O}(\sigma^2) \ .\eqnum{B.9}
\ee
A calculation similar to that in Eq.~(B.6) yields
\be
\frac{d}{dt}\bchi_t(\zeta) = X_t^* \biggl( \frac{d}{dt}
\chi_t(x) \biggr)\biggr|_{x=X_t(\zeta)} + \frac{d}{d\sigma} \biggl(
X^*_{t+\sigma} \chi_t(X_{t+\sigma}(\zeta))\biggr) \biggr|_{\sigma=0}
\eqnum{B.10}
\ee
for the fluid space covector
$\bchi_t(\zeta) := X^*_t \chi_t(X_t(\zeta))$.
Therefore the Lie derivative of $\chi_t$ is
\be
\Bigl(\pounds_{V_t} \chi_t\Bigr)\biggr|_{x} =
- \frac{d}{dt} \chi_t(x) + Z^*_t \biggl( \frac{d}{dt} \bchi_t(\zeta)
\biggr) \biggr|_{\zeta=Z_t(x)} \ .\eqnum{B.11}
\ee
Likewise, the Lie derivative of a contravariant vector is given by
\be
\Bigl(\pounds_{V_t} \chi_t\Bigr)\biggr|_{x} =
- \frac{d}{dt} \chi_t(x) + X^\prime_t \biggl( \frac{d}{dt}
\bchi_t(\zeta)  \biggr) \biggr|_{\zeta=Z_t(x)} \ ,\eqnum{B.12}
\ee
where $X^\prime_t$ is the derivative mapping of contravariant vectors.

Armed with the results (B.7), (B.11), (B.12) and their generalizations
to arbitrary tensors, Eq.~(3.12) is obtained as follows. For
definiteness, consider the case in which $\chi_t(x)$ is a one--parameter
family of covariant vectors and $\pi_t(x)$ denotes the canonically
conjugate one--parameter family of contravariant
vector densities on $\Sigma$. Contract Eq.~(B.11) with $\pi_t(x)$ and
integrate over $\Sigma$ to obtain
\bea
\int_\Sigma d^3x \,\pi_t(x) \frac{d}{dt} \chi_t(x) & = &
\int_\Sigma d^3x  \biggl( \pi_t(x) Z^*_t \biggl( \frac{d}{dt}
\bchi_t(\zeta) \biggr) \biggr|_{\zeta=Z_t(x)} - \pi_t(x)
\Bigl(\pounds_{V_t} \chi_t\Bigr)\biggr|_{x} \biggr) \nonumber\\
& = & \int_\Sigma d^3x \biggl( \pi_t(x) Z^*_t \biggl( \frac{d}{dt}
\bchi_t(\zeta) \biggr) \biggr|_{\zeta=Z_t(x)} -
V^a_t(x) H_a(x;\pi_t,\chi_t] \biggr) \ .\eqnum{B.13}
\eea
Here, $H_a(x;\pi_t,\chi_t]$ is the contribution from $\chi_t$ and
$\pi_t$ to the momentum constraint. (Integration by parts is used to
obtain the second line of Eq.~(B.13) from the first. It is assumed
that the space manifold $\Sigma$ is compact without boundary, so that
no boundary terms appear.) Now rewrite the right--hand side of
Eq.~(B.13) as an integral over the fluid space $\S$ by
changing integration variables, $x^a = X^a_t(\zeta)$. This gives
\be
\int_\Sigma d^3x \,\pi_t(x) \frac{d}{dt} \chi_t(x) =
\int_\S d^3\zeta \biggl( \bpi_t(\zeta) \frac{d}{dt} \bchi_t(\zeta)
- \boldV_t^k(\zeta) \boldH_k(\zeta;\bpi_t,\bchi_t] \biggr)
\ .\eqnum{B.14}
\ee
The vector field $\boldV_t(\zeta)$ is defined in terms of local
coordinates by
\be
\boldV_t^k(\zeta) := \biggl( V_t^a(x) {Z_t^k}_{,a}(x) \biggr)
\Bigr|_{x=X_t(\zeta)}  = \biggl( \frac{d}{dt} X^a_t(\zeta) \biggr)
\biggl( {Z^k_t}_{,a}(x) \biggr) \Biggr|_{x=X_t(\zeta)}
\ ,\eqnum{B.15}
\ee
where the definition (B.1) has been used. Inserting this expression for
$\boldV_t(\zeta)$ into Eq.~(B.14) yields Eq.~(3.12) from the main text.
The derivation for tensors of arbitrary rank follows along similar lines.


\begin{references}
\bibitem{Taub} See for example A. H. Taub, Phys. Rev. {\bf 94}, 1468 (1954);
  Commun. Math. Phys. {\bf 15}, 235 (1969).
\bibitem{HandE} See for example S. W. Hawking and G. F. R. Ellis, {\em
  The Large Scale Structure of Space--Time} (Cambridge University Press,
  Cambridge, 1973).
\bibitem{Lanczos} C. Lanczos, {\em The Variational Principles of Mechanics}
   (University of Toronto Press, Toronto, 1970).
\bibitem{Clebsch} A. Clebsch, J. Reine Angew. Math. {\bf 56}, 1 (1859).
\bibitem{TandO'H} K. Tam and J. O'Hanlon, Nuovo Cimento {\bf 62}B,
   351 (1969).
\bibitem{Schutz} B. F. Schutz, Phys. Rev. D {\bf 2}, 2762 (1970).
\bibitem{Sactions} B. F. Schutz, Phys. Rev. D {\bf 4}, 3559 (1971);
  J. R. Ray, J. Math. Phys. {\bf 13}, 1451 (1972);
  B. F. Schutz and R. Sorkin, Ann. Phys. {\bf 107}, 1 (1977);
  J. Demaret and V. Moncrief, Phys. Rev. D {\bf 21}, 2785 (1980);
  L. Bombelli and R. J. Torrence, Class. Quantum Grav. {\bf 7}, 1747 (1990).
\bibitem{Car} C. Carath\`{e}odory, {\em Variationsrechnung
  und partielle Differentialgleichungen erster Ordnung I} (Teubner,
  Leipzig, 1956).
\bibitem{Brown} J. D. Brown, Class. Quantum Grav. {\bf 10}, 1579 (1993).
\bibitem{TaubV} A. H. Taub, Arch. Ratl. Mech. Anal. {\bf 3}, 312 (1959).
\bibitem{Carter} B. Carter, Commun. Math. Phys. {\bf 30}, 261 (1973).
\bibitem{Kijowski} J. Kijowski and W. M. Tulczyjew, {\em A Symplectic
  Framework for Field Theories} (Springer, Berlin, 1979);
  J. Kijowski, A. Sm\'{o}lski, and A. G\'{o}rnicka,
  Phys. Rev. D {\bf 41}, 1875 (1990).
\bibitem{Sbaractions} H. P. K\"{u}nzle and J. M. Nester, J. Math. Phys.
  {\bf 25}, 1009 (1984); B. Carter, in {\em Relativistic Fluid Dynamics},
  edited by A. Anile and Y. Choquet-Bruhat (Springer-Verlag, Berlin, 1989);
  G. L. Comer and D. Langlois, Class. Quantum
  Grav. {\bf 10}, 2317 (1993).
\bibitem{dust} J. D. Brown and K. V. Kucha\v{r}, ``Dust as a standard of
  space and time in canonical quantum gravity", submitted to Phys. Rev. D.
\bibitem{clocks} J. D. Brown and D. Marolf, ``Fluids and clocks in
  canonical quantum gravity", manuscript in preparation.
\bibitem{DeWitt} B. S. DeWitt, in {\em Gravitation: An Introduction to
  Current Research}, edited by L. Witten (Wiley, New York, 1962).
\bibitem{Dantzig} D. van Dantzig, Physica {\bf 6}, 693 (1939).
\end{references}
\end{document}